\documentclass[aps,epsf,floats,pre,nofootinbib]{revtex4-1}
\usepackage{graphics,graphicx,epsfig}
\usepackage{amssymb,color}
\usepackage{epsf,epstopdf,wrapfig}
\usepackage{amsmath}
\usepackage{multirow}
\def\be{\begin{equation}}
\def\ee{\end{equation}}
\def\beq{\begin{equation}}
\def\eeq{\end{equation}}
\def\bea{\begin{eqnarray}}
\def\eea{\end{eqnarray}}

\newcommand{\eref}[1]{Eq.~(\ref{#1})}%
\newcommand{\fref}[1]{Fig.~\ref{#1}} %
\newcommand{\aref}[1]{\hyperref[#1]{Appendix~\ref{#1}}}

\newcommand{\lx}{\left}
\newcommand{\rx}{\right}

\newcommand{\sumi}  {\sum_i}
\newcommand{\sumj}  {\sum_j}
\newcommand{\sumij} {\sum_{i,j}}

\newcommand{\nmax}  {n_{\mathrm{max}}}

\newcommand{\summuz} {\sum_{\mu=0}^M}
\newcommand{\summuo} {\sum_{\mu=1}^M}

\newcommand{\sumpn} {\sum_n{}^{'}}

\newcommand{\si}{\vec{s_i}}
\newcommand{\sj}{\vec{s_j}}

\newcommand{\kij}{k_{ij}}
\newcommand{\kji}{k_{ji}}

\newcommand{\hdkimmn}{\delta(\kim - n)}
\newcommand{\hdkmimn}{\delta(\kmi - n)}

\newcommand{\hdklmmn}{\delta(\klm - n)}
\newcommand{\hdkmlmn}{\delta(\kml - n)}
\newcommand{\hdkijmn}{\delta(\kij - n)}
\newcommand{\hdkjimn}{\delta(\kji - n)}

\newcommand{\klm}{k_{lm}}
\newcommand{\kml}{k_{ml}}
\newcommand{\kim}{k_{im}}
\newcommand{\kmi}{k_{mi}}

\newcommand{\costij}{\cos\lx(\theta_{ij}\rx)}
\newcommand{\ThetaL}{\Theta\lx(\lx|\costij\rx| - 1/2\rx)}
\newcommand{\ThetaT}{\Theta\lx(1/2 - \lx|\costij\rx|\rx)}

\newcommand{\StdHmlt}{\sumij \Jij \si \cdot \sj}

\newcommand{\piv}[1]{\vec{\pi}_{#1}}
\newcommand{\pipv}[1]{\vec{\pi^{\prime}}_{#1}}
\newcommand{\pip}{\pi^{\prime}}

\newcommand{\avefexp}[1]{\langle{f_{#1}\rangle_\mathrm{exp}}}
\newcommand{\avefens}[1]{\langle f_{#1} \rangle_\mathrm{P}}
\newcommand{\Jij}{J_{ij}}
\newcommand{\CorrN}{\hat{C}(n)}
\newcommand{\CorrNL}{\hat{C}^L(n)}
\newcommand{\CorrNT}{\hat{C}^T(n)}

\newcommand{\CorrNexp}{\langle \hat{C}(n) \rangle_\mathrm{exp}}
\newcommand{\CorrNLexp}{\langle \hat{C}^L(n) \rangle_\mathrm{exp}}
\newcommand{\CorrNTexp}{\langle \hat{C}^T(n) \rangle_\mathrm{exp}}
\newcommand{\CorrNLTexp}{\langle \hat{C}^{L,T}(n) \rangle_\mathrm{exp}}

\begin{document}

\title{Short range interactions vs long range correlations in bird flocks}

\author{Andrea Cavagna$^{a,b,c}$, Lorenzo Del Castello$^{a,b}$, Supravat Dey$^{a,b}$, Irene Giardina$^{a,b,c}$,  Stefania Melillo$^{a,b}$,  \\ Leonardo Parisi$^{a,b,d}$, and Massimiliano Viale$^{a,b}$}

\affiliation{$^a$ Istituto Sistemi Complessi, Consiglio Nazionale delle Ricerche, UOS Sapienza, 00185 Rome, Italy}
\affiliation{$^b$ Dipartimento di Fisica, Universit\`a\ Sapienza, 00185 Rome, Italy}
\affiliation{$^c$ Initiative for the Theoretical Sciences, The Graduate Center, 365 Fifth Avenue, New York, NY 10016 USA}
\affiliation{$^d$ Dipartimento di Informatica, Universit\`a\ Sapienza, 00198 Rome, Italy}


\begin{abstract}
Bird flocks are a paradigmatic example of collective motion. One of the prominent experimental traits discovered about flocks is the presence of long range velocity correlations between individuals, which allow them to influence each other over the large scales, keeping a high level of group coordination. A crucial question is to understand what is the mutual interaction between birds generating such nontrivial correlations. Here we use the Maximum Entropy (ME) approach to infer from experimental data of natural flocks the effective interactions between birds. Compared to previous studies, we make a significant step forward as we retrieve the full functional dependence of the interaction on distance and find that it decays exponentially over a range of a few individuals. The fact that ME gives a short-range interaction even though its experimental input is the long-range  correlation function, shows that the method is able to discriminate the relevant information encoded in such correlations and single out a minimal number of effective parameters. Finally, we  show how the method can be used to capture the degree of anisotropy of mutual interactions.

\end{abstract}   


\maketitle

\section{Introduction}

Large groups of animals - such as bird flocks, fish schools and insect swarms - display a remarkable degree of collective coordination. Several experimental studies in the last decade quantified the spontaneous emergence of global order \cite{Camazine_book03,Couzin03,vicsek12}, the presence of strong behavioral correlations between individuals \cite{Cavagna10,Attanasi13c}, and the swift transfer of information through the group \cite{Couzin03, Sumpter08, Attanasi13a, Sumpter14}. Such findings stimulated  a multi-disciplinary interest in this kind of systems. On the one hand, animal groups can be considered as instances of active matter \cite{rama10,review_active}, and can be expected to display some of the non trivial properties observed and predicted for several living, soft and granular active systems on the micro-scale. On the other hand, there are important features that make animal groups more complicated to understand. First, the way individuals coordinate with one another are not only determined by physical mechanisms, as for rods or hard discs, but also (and often mainly) by exquisitely biological processes (including cognitive). As a consequence, any speculation about the nature of mutual interactions in a group cannot be taken for granted. Besides, animal aggregations form large, but not infinitely large groups: they are not in the thermodynamic limit, but rather live in an intermediate regime where finite size effects can be important \cite{Attanasi13c}.
Understanding collective animal behavior therefore implies understanding what is the nature of interactions, what are the effective features of such interactions that are relevant on the scale of natural groups, and how they determine the collective properties that we observe.

One of the most intriguing features of collective animal motion is the presence of long-range correlations. The correlation function of the velocity fluctuations has been found to be long-range  both in polarized groups as bird flocks \cite{Cavagna10} and in disordered ones, such as insect swarms \cite{Attanasi13c}. These results suggest that, rather than ordering, what is truly  characteristic of collective behavior in biological systems is the ability of individuals to correlate changes in behavior and influence each other over the large scales.  It is therefore important to understand what are the features of the interactions granting such strong correlations.

We know from statistical physics that short-range interactions are sufficient to produce spontaneous symmetry breaking and system-level coordination. Models of self-propelled particles \cite{Vicsek95,chate04,chate08,ginelli10,Charlotte12,vicsek12} and hydrodynamic flocking theories \cite{Toner95,rama10} have shown numerically and analytically that in active systems also short-range interactions can produce global ordering and long-range correlations.  Indeed, there now seems to be some consensus in the field of collective animal behavior that interactions are short-ranged \cite{Sumpter_book10}. 
However, long-range interactions do exist in nature, so we cannot rule them out {\it a priori}. Moreover, to create long-range correlations out of short-range interactions one normally needs some special conditions: either there is a continuous spontaneously broken symmetry (Goldstone theorem), or the system is in the scaling region of a critical point. From a biological perspective, one could legitimately object that a reasonable long-range interaction is a better explanation of long-range correlation than some arcane physical theorem, not to mention criticality.  For example, birds' vision is likely to span the entire size of a flock. 
It is worth noticing that, despite the short-range consensus, it has been recently proposed that a {\it long-range} interaction is at the basis of flocking behavior \cite{Turner14}. Hence, the notion that short-range interaction rules collective behavior, albeit reasonable, is still far from being an established fact, even in those system that have been most studied experimentally.

Although the access to large scale empirical data on animal groups in the last decade or so has considerably advanced our understanding of the problem, we still have no direct and unbiased proof of whether interaction is short- or long-ranged. Significant results have been obtained by fitting biological models to the data \cite{Lukeman10,Gautrais12,Couzin11,Ouellette14}. Yet the problem with model fitting is that it may be tricky to distinguish the intrinsic properties of the system under investigation from the a priori ingredients of the model used to fit the data. Alternatively, interaction has been assessed by using some structural proxy of it; for example, in \cite{Ballerini08a} it has been found experimentally that the anisotropy in the nearest neighbours spatial distribution decays in a short-range way. However, through this kind of structural proxies one does not attain {\it direct} access to the interaction.

In this paper we follow a different approach  and use the maximum entropy method \cite{Jaynes57} to infer the interactions directly from the data. The philosophy of this method is different from standard model fitting in that, as we will discuss, the model it designs for the system is dictated by the available experimental observables and it is not assumed  a priori  For bird flocks, we started this program in \cite{Bialek12}, with encouraging results. Using  a very simple experimental input, we inferred the effective number of individuals each bird is interacting with, and the average strength of such interaction.
Hence, in \cite{Bialek12} it was {\it assumed} a step-like shape of the interaction, in order to keep the mathematical complexity to a minimum.
Here, we make a significant step forward and derive the {\it full functional dependence on distance}
 of the effective alignment interaction between individuals. We call this function $J(n)$, where $n$ is the topological distance between birds, i.e. their 
 order of neighborhood \cite{Ballerini08a}.
   We find that $J(n)$ decays exponentially, on scales much smaller than the system size, hence indicating that alignment interaction within a flock is short-range. The experimental input of our calculations is the  velocity correlation function, which is long-range. We show however that much of the information captured by the correlation function is redundant and only correlations on a short scale are sufficient to retrieve the interaction $J(n)$.   Hence, not only can we infer the effective interaction, but we only need a small number of local experimental measurements. Thanks to the new method we are also able to study the angular dependence of the interaction with respect to the direction of motion of the flock and shed some light on the anisotropic spatial arrangement of neighboring birds found in past experimental studies \cite{Ballerini08a,Ballerini08b}. 

The paper is organized in the following way. In Sec.\ref{sec:maxent} we introduce and describe the Maximum Entropy approach, we outline the mathematical structure of the computation and apply it to the case of flocks. In Sec.\ref{sec:results} we show the results of the computation for the flocking events in our dataset; we compare them with previous work \cite{Bialek12}; and we further generalize the method to capture the possible angular anisotropy of the interactions.
In Sec.\ref{sec:nmax} we discuss what are the effects of changing the number of experimental input parameters in the calculation and show that a fair result is achieved when the inferred interaction does not depend on the number of input parameters anymore. Finally, in Sec.\ref{sec:conclusions} we discuss our results.

\section{Maximum Entropy approach to flocks}
\label{sec:maxent}

Collective phenomena and ordering transitions have been widely studied in condensed matter. From the perspective of statistical physics, one usually knows the microscopic interactions between particles, and wants to predict their large scale properties. When dealing with biological systems we often face the opposite situation. We have access through experiments to collective observables, but have scarce knowledge on the effective interactions generating them. The problem is in this case an {\it inverse} one: to build a microscopic statistical model starting from the macroscopic data. As mentioned in the Introduction, several approaches have been developed to deal with this task, from model fitting to Bayesian inference \cite{Mackay}. Here we consider the Maximum Entropy (ME) approach. This method was originally established by E. T. Jaynes in 1957 \cite{Jaynes57} and has strong connections with classical statistical physics. In the last decade, it has been widely used to describe the collective behavior of biological networks, from neural assemblies, to amino acids in proteins, biochemical and genetic networks and flocks of birds \cite{Schneidman06,Shlens06,TkacikReview10,Tkacik13,Tkacik14,Halabi09,Weigt09,Mora10,Lezon06,Bialek12,Bialek14,cavagna14,Castellana15}. 

The main idea of the ME method is to build the least structured statistical model - the maximum entropy model - which is consistent with a given set of measured observables. In this section we explain how to construct a ME model and how the method can be applied to the case of birds flocks. Before doing this, however, we would like to make two remarks on the method, which are useful to understand its philosophy, appreciate its results and evaluate its performance.

i) As compared to other approaches, the ME principle has the remarkable feature that it does not rely on a priori assumptions on the system under study; this means that ME does not assume any form of the microscopic interactions (at variance with  model fitting). This does not mean that the ME does not make approximations in the description of the system; in fact it does, but we have a way to control and evaluate them.  
As we shall see in detail in this Section, the kind of model we get from the ME approach crucially depends on the experimental observables we consider as input. Were we able to perform good measurements of many observables we would retrieve in an accurate way the full probability distribution of the micro-states of our system. This is not however what happens in real experiments, where typically only a few quantities can be measured, and not always  in a robust statistical way. What we know is that, given some observables, the ME model is the one that describes them best with the least number of assumptions. In this sense, the effective interactions appearing in a ME model only come from the experimental behavior of the system. Besides, once we construct a ME model based on some experimental input, we have a way to test its predictive power.  We can, for example, use the model to predict quantities other than the ones used to build it, and compare to experiments. More systematically, we can compute the predictive gain acquired when providing new experimental input and assess the information content of the ME model. What happens in some cases is that a few experimental input observables give significant gains, and adding further experimental input makes weak progress.  We will discuss an example of this procedure in Sec.\ref{sec:nmax}.

ii) If we consider as input observables quantities that are - at least on certain timescales - stationary, the ME method provides a static ME model. As we shall see, in this case the ME distribution has the form of a Boltzmann  measure, which is particularly useful from a mathematical point of view to perform computations. This does not mean, however, that the system is in equilibrium, nor that the ME distribution is an equilibrium one. In fact the system can have an arbitrary off-equilibrium dynamics. In this case the ME method simply captures the effect of this dynamics on the statistical distribution of  a given set of observables. What is relevant to us in this paper is that this distribution encodes how certain degrees of freedom effectively interact due to the microscopic behavior of the system. We note that the ME approach is not bound to produce static Boltzmann-like measures. If we consider as input observables time dependent quantities (such as multi-point time correlation functions), the resulting ME model will consist in a time dependent distribution \cite{vander09,marre09,roudi11,cavagna14}. Computations and inference of effective interactions can be in this case much more complicated. For polarized self-propelled systems we showed that, as long as the network of positions does not rearrange too fast (which is the case of natural flocks \cite{diffusion}), static and dynamic ME models give a very similar inference of the interaction parameters \cite{cavagna14}.

\subsection{The general ME scheme}
\label{sec:maxentgen}

Consider a system whose micro-state at any instant of time is described by a set of variables $\{{\bf x}_1, {\bf x}_2, ..., {\bf x}_i, ...,{\bf x}_N\} \equiv {\bf X}$. When the size of the system $N$ becomes large, the space of the ${\bf X}$ increases exponentially and it is experimentally impossible to directly sample and reconstruct the probability distribution $P({\bf X})$. On the contrary, it is usually possible to accurately measure aggregate observables, which require less statistics. 
Let us assume that we can measure several observables $f_1({\bf X}), f_2({\bf X}), ... f_M({\bf X})$, and let us denote their experimental averages by $\avefexp{1}, \avefexp{2}, ..., \avefexp{M}$, respectively. The maximum entropy (ME) method consists in finding the most random probability distribution $P({\bf X})$ that is consistent with the observed experimental data. The distribution must therefore satisfy the following constraint:
\bea
 \avefexp{\mu}
= \avefens{\mu}  \ ,
\label{eqn:constraint0}
\eea
for all $\mu=1, 2,..., M$, and where $\avefens{\mu}=\sum_{\bf X} P({\bf X})f_{\mu}({\bf X})$ denotes the expectation value computed using the probability distribution $P({\bf X})$. Many distributions satisfy \eref{eqn:constraint0}. The maximum entropy principle \cite{Jaynes57} aims to find the one which has as little structure as possible, i.e. is the most random, so that one can derive the minimal consequences of the experimental observations on $\avefexp{\mu}$. As a measure of randomness of a given distribution we consider its Shannon entropy \cite{Shannon48,Cover},
\bea
S[P] = -\sum_{\bf X} P({\bf X})\log{P({\bf X})}.
\label{eqn:entropy}
\eea    
In order to get the desired probability distribution, we then need to maximize $S[P]$ under the constraints given by \eref{eqn:constraint0}. Apart from the experimental constraints, there is an additional constraint, namely that the probability distribution should be normalized $\sum_{\bf X} P({\bf X}) = 1$. This is equivalent to say that we add to our list of observables an extra function, the constant $f_{0}({\bf X}) = 1$.
This constraint maximization problem can be solved with the Lagrange multiplier method \cite{Bender_SM} by finding the optimum of the generalized entropy function,
\bea
S[P;\{\lambda_{\nu}\}] &=& S[P] - \summuz \lambda_{\mu} \lx( \avefens{\mu} - \avefexp{\mu} \rx),
\label{eqn:lagrange}
\eea   
where each Lagrange multiplier $\lambda_{\mu}$ is associated with a constraint equation. Maximizing $S[P;\{\lambda_{\nu}\}]$ with respect to $P({\bf X})$, we get  
\bea
P({\bf X}) &=& \frac{1}{Z({\{\lambda_{\nu}\}})} 
\exp\left[ -\summuo \lambda_{\mu} f_{\mu}({\bf X})\right],
\label{boltz}
\eea
where $Z(\{\lambda_{\nu}\})$ enforces the normalization and is obtained optimizing with respect $\lambda_0$ 
\be
Z(\{\lambda_{\nu}\}) =\exp(1 + \lambda_0)
 = \sum_{\bf X} \exp\lx[ -\summuo \lambda_{\mu} f_\mu({\bf X})\rx].
\label{eqn:partfunc}
\ee
Using Eq.(\ref{boltz}) the generalised entropy (\ref{eqn:lagrange}) can be written as a function of the Lagrange parameters only, 
\bea
S(\{\lambda_{\nu}\}) &=& \log{Z(\{\lambda_\nu\})} + \summuo \lambda_{\mu} \avefexp{\mu}.
\label{eqn:entropy2}
\eea
One can now easily optimize with respect to $\lambda_{\mu}$ to recover the original constraint equation (\ref{eqn:constraint0}),
\bea
-\frac{\partial \log Z({\{\lambda_{\nu}\}})}{\partial \lambda_\mu} &=& \sum_{\bf X} P({\bf X}) f_\mu({\bf X}) = \avefexp{\mu}.
\label{eqn:derivativeeqn}
\eea
As we can see from Eq.(\ref{boltz}), the maximum entropy distribution has the form of a Boltzmann distribution $P({\bf X})=\exp(-\beta H({\bf X}))/Z$ with an effective ``Hamiltonian'' $H({\bf X})=\summuo \lambda_\mu f_\mu({\bf X})$ and temperature $k_BT=1$. As we previously discussed, this does not mean that the system we are looking at is in equilibrium nor that this Hamiltonian is the true microscopic Hamiltonian of the system (if it exists). Nevertheless, one must not forget that the optimal values of the Lagrange parameters, through Eq.(\ref{eqn:derivativeeqn}), enforce consistency with experimental data. They describe the effect of the microscopic dynamics on the statistics of the input observables. In this sense, they represent effective interactions and mirror the structure of the microscopic behavior of the system through the filter of our experiments. In this respect, the choice of which experimental observables to consider as input of the method is very important: the more representative the set of $\{f_\mu({\bf X})\}$ is of the collective behavior of the system, the more predictive the ME model (\ref{boltz}) turns out to be, the more informative the effective parameters are on the microscopic features of the system that determine its behavior at the collective scale. 

Keeping these considerations in mind, to investigate the nature of interactions in our system, we need to select a good set of experimental input observables, compute the corresponding ME model, investigate its predictive content, and finally look at the structure of the effective ME interactions. 

From a mathematical point of view, to compute the ME model we need to solve Eqs. (\ref{boltz}) and (\ref{eqn:derivativeeqn}). This means computing  $Z(\{\lambda_{\nu}\})$, which represents the partition function of the Boltzmann-like distribution (\ref{boltz}), a problem for solving which we are fairly well-equipped (at the level of schemes and approximations) in statistical physics. There is, however, a further obvious difficulty: $Z$ is not a number, but a function of the Lagrange parameters, and must be computed for any possible value of the $\{\lambda_{\nu}\}$. This is the essence of the inverseproblem. In most cases this is a hard step, which is achieved numerically. For flocks, which are very polarized groups, one can resort to a high-order expansion and compute $Z(\{\lambda_{\nu}\})$ analytically. Once this function is known, we can fix the values of the Lagrange parameters by enforcing the constraint, i.e. by optimizing Eq.(\ref{eqn:derivativeeqn}). 

Interestingly, we note that the generalized entropy (\ref{eqn:derivativeeqn}) is related to the likelihood of the experimental data  ${\mathcal{L(\{\lambda_{\nu}\})}}$, i.e.
\bea
\nonumber
\log{\mathcal{L(\{\lambda_{\nu}\})}} &=& \langle \log{P} \rangle_{\mathrm{exp}}
=-\log Z(\{\lambda_\nu\}) - \summuo \lambda_{\mu} \avefexp{\mu} =  - S(\{\lambda_{\nu}\})
\label{eqn:loglikelihood}
\eea
Hence, optimizing the generalized entropy (which - as can be shown - corresponds to a minimum in the parameters' space) is equivalent to maximize the log-likelihood of the experimental data.

\subsection{ME distribution for flocks}
\label{sec:maxentflock}

Let us now apply the ME method to flocks of birds. The first step is to identify the set of variables that
defines the microstate of the system under investigation (i.e. the variables $\{{\bf X}\}$ of the previous section). We are
interested in the interaction that is responsible for the  alignment of the directions of motion of the birds. 
Hence we consider as microscopic variables the orientation vectors,  $\vec{s}_i\equiv \vec{v}_i/|\vec{v}_i|$,  where $\vec v_i$ is the velocity of bird $i=1,\dots,N$. 

The second step is to select a set of observables, function of the variables  $\vec{s}_i$, whose 
experimental value will be used to constrain the probability distribution of the orientation vectors, $P(\{\vec s_i\})$. 
The standard way to proceed is to use moments (i.e. correlations) of this distribution, $\langle \vec s \rangle, 
\langle \vec s \cdot \vec s \rangle, \dots$. As one can easily see from Eq.(\ref{boltz}), each one of these $m$-points correlations will 
generate $m$-points interaction terms in the effective Hamiltonian. One could naively think that the more correlations we consider, the better the corresponding ME model. In fact this not true, for several reasons. On the experimental side, the larger is $m$ the larger is the statistics needed to get good experimental estimates (in terms of number of measurements and sample size). Thus,  using large $m$-point correlations typically enhances the experimental noise. From a more conceptual point of view, not all correlations are in general equally important. By considering too many of them we can introduce redundant information and risk to overfit the parameters. The most economic prescription is therefore to use up to the minimum $m$-point correlation that allows to predict the $m+1$-point correlation.

Previous studies have shown that in flocks (as in other collective systems \cite{Schneidman06}) the use of
pairwise interactions ($m=2$) allows to accurately predict the $4$-points correlations \cite{Bialek12,Bialek14}. We then  focus on pairwise correlations. In a flock of birds we can in principle define the mutual correlation of the flight directions $C_{ij}=\vec{s}_i\cdot\vec{s}_j$ for any single pair of individuals. However,  these quantities wildly fluctuate in time and never reach a steady state. The reason is obvious: birds are not on a fixed lattice; they move in space so to change position with respect to each other. Therefore, the mutual distance of $i$ vs $j$ changes in time; but any reasonable social force (i.e. interaction) will depend on the {\it relative} distance between individuals, rather than on their {\it absolute} identity. Hence, distance, rather than identity, should be used as a label.
 The experimental proof of that is that  the correlations computed as a function of distance are stable in time over appreciable intervals.  Besides, they exhibit a non-trivial scale free dependence, signature of the collective behavior of the flock \cite{Cavagna10}, which makes them a very good choice for our purposes.

 We therefore consider as our experimental input observable the two-point correlation function, 
\bea
\hat C(n; \{\vec s_i\}) = \frac{1}{N} \sum_{i,j=1}^N \vec{s}_i\cdot\vec{s}_j \;
\delta(k_{ij}-n) \ .
\label{eqn:cn}
\eea
This quantity is the average correlation between a bird and its $n^\mathrm{th}$ nearest neighbour (we use the hat notation
to distinguish this full correlation from the connected one - see below).
Compared to previous studies \cite{Cavagna10}, we measure the correlation as  a function of the topological distance (i.e. order of neighborhood), $n$, rather than of the metric distance, $r$ \cite{Ballerini08a}. In Eq.(\ref{eqn:cn}) $k_{ij}$ is the topological distance of bird $j$ relative to bird $i$: if $j$ is the first nearest neighbour of $i$,  $k_{ij}=1$; if $j$ is the second nearest neighbour of $i$,  $k_{ij}=2$; and so on ($k_{ij}$ is nonsymmetric).  This implies $\sum_{i,j} \delta(k_{ij}-n)=N$, and this is why the normalization in (\ref{eqn:cn}) is easier than in its metric counterpart \cite{Cavagna10, footnote1}.

As explained in the previous section, the ME method consists in finding the probability distribution $P(\{\vec s_i\})$ that maximizes the entropy $S[P]$ under the constraint that the distribution reproduces the experimental observables (\ref{eqn:cn}), which in our case read
\begin{equation}
\langle \hat C(n; \{\vec s_i\}) \rangle_\mathrm{exp} 
=
\langle \hat C(n; \{\vec s_i\}) \rangle_\mathrm{P} 
 \ .
\label{suka0}
\end{equation}
This constrained maximization is achieved by introducing one Lagrange multiplier, $J(n)$, for each 
experimental quantity that we are fixing, $\hat C(n)$ (which represent, respectively, the $\lambda_\mu$ and $f_\mu$ of sec.\ref{sec:maxentgen}). As we have explained,  the distribution obtained in this way has the form of an exponential of the product of the  Lagrange multipliers times the observables 
\bea
P(\{\vec s_i\}) = \frac{1}{Z} \ e^{ N \sum_n J(n) \hat C(n)} = 
\frac{1}{Z}\ e^{ \sum_{ij} J(k_{ij})  \,\vec{s}_i\cdot\vec{s}_j   }
 \ ,
\label{eqn:model1}
\eea
where $Z$ is the normalizing partition function and $J(k_{ij})=\sum_n J(n) \delta(n-k_{ij})$.  The probability distribution (\ref{eqn:model1}) corresponds to the effective Hamiltonian, 
\bea
H=-\sum_{ij} J(k_{ij})  \,\vec{s}_i\cdot\vec{s}_j \ .
\label{eqn:model1H}
\eea
Therefore, the (discrete) function $J(n)$ represents the strength of the effective alignment interaction between pairs of birds at topological distance $n$.  Once we solve the ME model and we compute $J(n)$, we can therefore investigate the nature of such interaction, how it decays in distance and understand whether it is short- or long-ranged.

Inferring the full function $J(n)$ is a significant step forward compared to our previous ME calculations \cite{Bialek12, cavagna14}, where
we assumed a step-like shape of the interaction. By doing that we only had to infer two parameters, intensity and range of the interaction, so
that we had no information about the {\it form} of the interaction. For this reason, the question of short vs long range interaction in \cite{Bialek12}
was addressed in a rather indirect way, namely by checking that the interaction range did not scale with the system size. Here, on the contrary, we will be able to calculate directly how the interaction decays and to see explicitly that it is short range.

\subsection{Maximization of log-likelihood}
\label{sec:likelihood}

To retrieve the interaction function $J(n)$ we need to solve the ME equations enforcing the constraints (\ref{suka0}) or, equivalently, maximize the log-likelihood of the data. In our case the log-likelihood function Eq.(\ref{eqn:loglikelihood}) is given by 
\begin{equation}
\log \mathcal{L} = \langle \log P(\{\vec s_i \}) \rangle_\mathrm{exp}  =-\log Z[J(n)] + N \sum_n J(n) \langle \hat C(n; \vec{s}_i)\rangle_\mathrm{exp}.
\label{eqn:loglikelihoodfull0} 
\end{equation}
We therefore need to compute the partition function $Z[J(n)]$ and then perform the maximization with respect to $J(n)$. This is a non-trivial program. There are however a few tricks we can exploit to facilitate the task. We outline here the main steps, details can be found in the Appendixes.
\begin{itemize}
\item[$\bullet$] The expression of the effective Hamiltonian can be simplified further. We can indeed rewrite  $H$ by introducing the symmetrized interactions matrix $J_{ij}$,
\begin{equation}
H=-\sum_{ij} J_{ij}  \,\vec{s}_i\cdot\vec{s}_j \,
\label{eqn:isotropicH}
\end{equation}
where $J_{ij} \equiv [{J}(\kij) + {J}(\kji)]/2$.
Interestingly,  Eq. (\ref{eqn:isotropicH}) describes the Hamiltonian of an Heisenberg model on a network, whose topology is described by the interaction matrix  $J_{ij}$. In the strongly ordered phase - as flocks are - this model can be solved using a well known low temperature expansion, the spin-wave approximation (see Appendix). As a result, $Z$ can be computed analytically and is entirely given in terms of the eigenvalues $\{a_k\}$ of discrete Laplacian matrix $A_{ij} = \delta_{ij} \sum_k J_{ik} -  (1 - \delta_{ij}) J_{ij}$, giving
\be
Z[J(n)]=-\sum_{k>1} \log{a_k}+\sum_n J(n)
\ee

\item[$\bullet$] In principle we need to consider all possible values of mutual distances $n=1\cdots N$, and  optimize over $N$ distinct Lagrange parameters. This number can be however severely reduced. It turns out (see next section) that correlations $C(n)$ for $n>n_{max}$ are redundant and do not improve the computation. Thus, all the sums appearing in Eq. (\ref{eqn:loglikelihoodfull0})  can be extended only up to $n_{max}$. Besides, one can 'bin' the integer values of topological distances $n$ in discrete intervals of size $\Delta n$, much as one would do with  real values of the metric distance (see Appendix \ref{app:coarse}). In this way the number of effective variational parameters can be reduced even further, speeding up the maximization procedure.  
The expression of the log-likelihood then becomes
\bea
\log{\mathcal{L}} &=&\; \sum_{k>1} \log{a_k} - N \Delta n  \sumpn J(n) (1 - \CorrNexp). \;
\label{eqn:loglikelihoodfull1}
\eea
where the primed sum indicates that we are summing over discrete bins, up to $n_{max}$.
\item[$\bullet$] The derivatives of the second term of the log-likelihood with respect to the $J(n)$ are trivial. However, differentiating the partition function is far less trivial, as the eigenvalues $a_k$ are very complicated functions of the $\{J(n)\}$. Luckily we can use perturbation theory (see Appendix \ref{app:der}) and derive the exact expressions for the derivatives w.r.t. to $J(n)$. 
\end{itemize}

In this way we finally get the ME equations
\bea
 \CorrNexp &=& 1-\frac{1}{N \Delta n} \sum_{k>1} \frac{1}{a_k} \frac{\partial a_k}{\partial J(n)} 
= 1-\frac{\text{Tr}[A^{-1}\gamma(n)]}{N \Delta n}.
\eea
where the matrix $\gamma$ is given by
\bea
\gamma_{ij}(n) &=& \frac{1}{2} \delta_{ij} \lx[ \sum_m (\hdkimmn + \hdkmimn) \rx] 
- \frac{1}{2} (1 - \delta_{ij}) (\hdkijmn + \hdkjimn).
\eea
These equations can be exploited to efficiently maximize the log-Likelihood (\ref{eqn:loglikelihoodfull1}) numerically (see Appendix \ref{app:num} for details on the numerical procedure) and find, for each value of $n$, the optimal $J(n)$. The results of this procedure are discussed in the next Section.

\section{Results}
\label{sec:results}

\subsection{Short range interactions vs long range correlations}
Let us summarize the procedure explained so far. We considered a set of experimentally measured observables, the velocity correlation functions eq.(\ref{eqn:cn}), and built the ME distribution consistent with these observables. This distribution is expressed in terms of effective alignment interactions between individuals, whose dependence in mutual distances is described by the function $J(n)$. The ME allows us to retrieve $J(n)$ by maximizing the log-likelihood, given the experimental input $ \CorrNexp$.

Let us now discuss the results of this procedure. We used an experimental dataset of 22 flocking events (see Table \ref{table} and Appendix \ref{app:data}). Data were obtained from stereoscopic experiments in the field: large flocks of starlings (from hundreds to thousands birds) were filmed with high resolution stereo-cameras and - thanks to innovative computer vision techniques - individual 3D tracking was performed \cite{Ballerini08a,Cavagna10,Attanasi13b}. Given the difficulty of the problem, this dataset represents to date the largest experimental dataset on large animal groups  moving in three dimensions.

\begin{figure}[]
\centering
\includegraphics[width=0.7\textwidth]{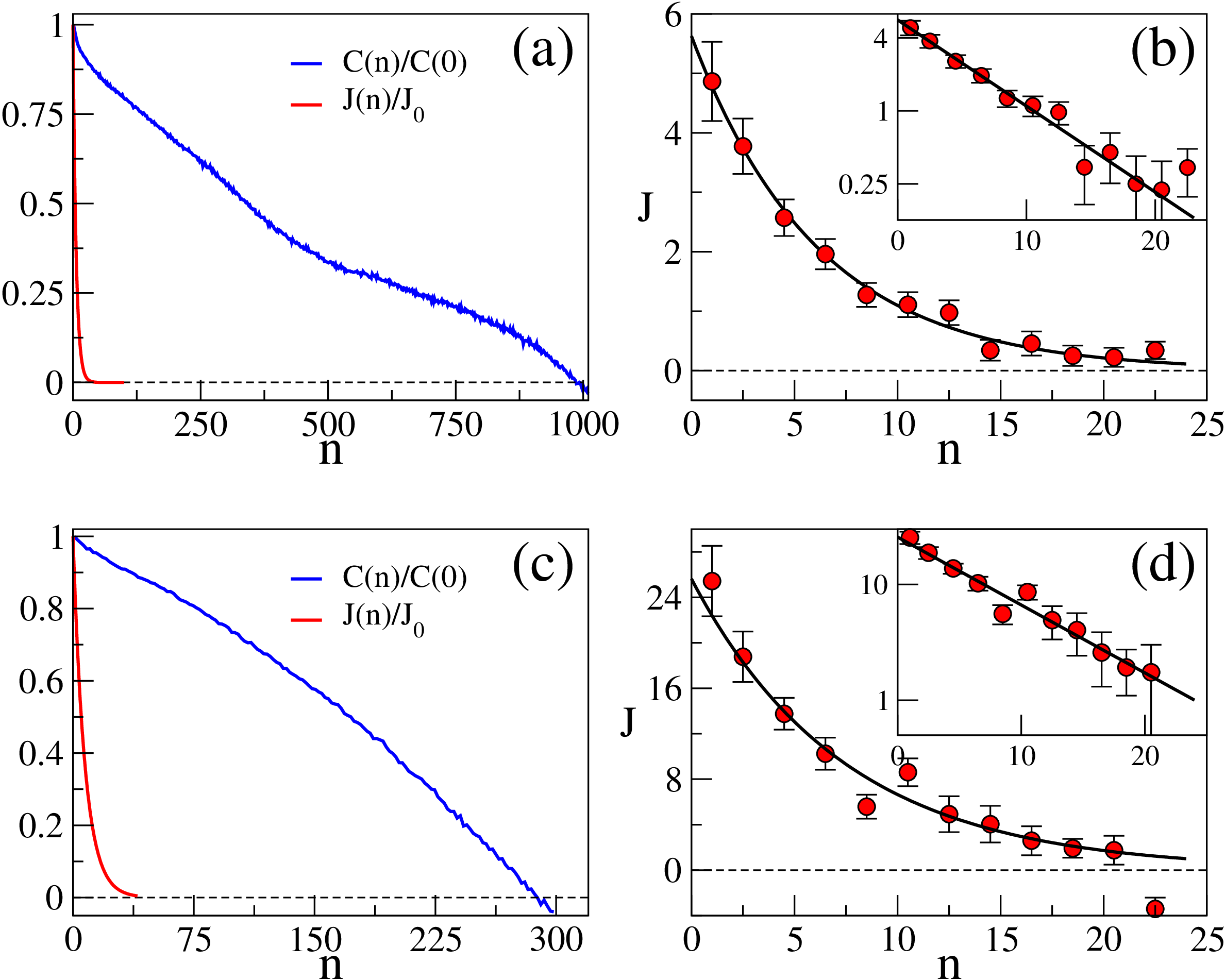}
\caption{
{\bf Left} 
Connected correlation function
$C(n)$ compared to the interaction $J(n)$ (both normalized by their $n=0$ value 
to show on the same scale).
{\bf Right} Close-up of the interaction, $J(n)$.
The full line is an exponential fit to the data (see text). 
Inset: Semi-log plot of the same quantity.
{\bf Top.} Event 31-01, $N=2126$; $J_0=5.63$, and $n_c=6.11$.
{\bf Bottom.} Event 21-06, $N=717$; $J_0=25.63$, and $n_c=7.41$.
}
\label{fig:fig1}
\end{figure}

For each event, we measured the correlations $ \CorrNexp$ and used them as input for the ME computation. 
The resulting $J(n)$  is plotted in Fig.\ref{fig:fig1}, for two distinct flocks. As we can see from the figure, the interaction function (red line) clearly decays to zero on a topological scale of few (order ten) individuals. To fully appreciate the result and its consequences we also plotted in the same figure the connected correlation function (blue line), which measures the decay of correlations between birds.  So far we always considered the {\it nonconnected} velocity correlation function  $\hat C(n)$, Eq.\ref{eqn:cn}. Flocks are in the ordered phase (they have nonzero polarization), hence $\hat C(n)$ does not decay to zero. This is simply a consequence of the emergence of long-range order. For this reason, if we want to describe how correlations decay, we need to consider the {\it connected} correlation function, which is defined by using the velocity fluctuations
\be
C(n) = \frac{1}{N}\sum_{i,j} \delta\vec{s}_i\cdot\delta\vec{s}_j \, \delta(k_{ij}-n) \ ,
\ee
where $\delta\vec{s}_i=\vec{s}_i -(1/N)\sum_k\vec{s}_k$. $C(n)$ is basically the full correlation minus the order parameter squared and, unlike $\hat C(n)$,  it does decay to zero for large distances. 
Yet, it was found in \cite{Cavagna10} that in starling flocks the connected correlation function is {\it long-ranged} (or scale-free), meaning that the correlation length, $\xi$, scales linearly with the system's size, $L$. In this respect, $C(n)$ describes the cooperative nature of collective motion in flocks and represents the non-trivial contribution to the experimental input of the ME calculation.

\begin{figure*}[t]
\centering
\includegraphics[scale=.35]{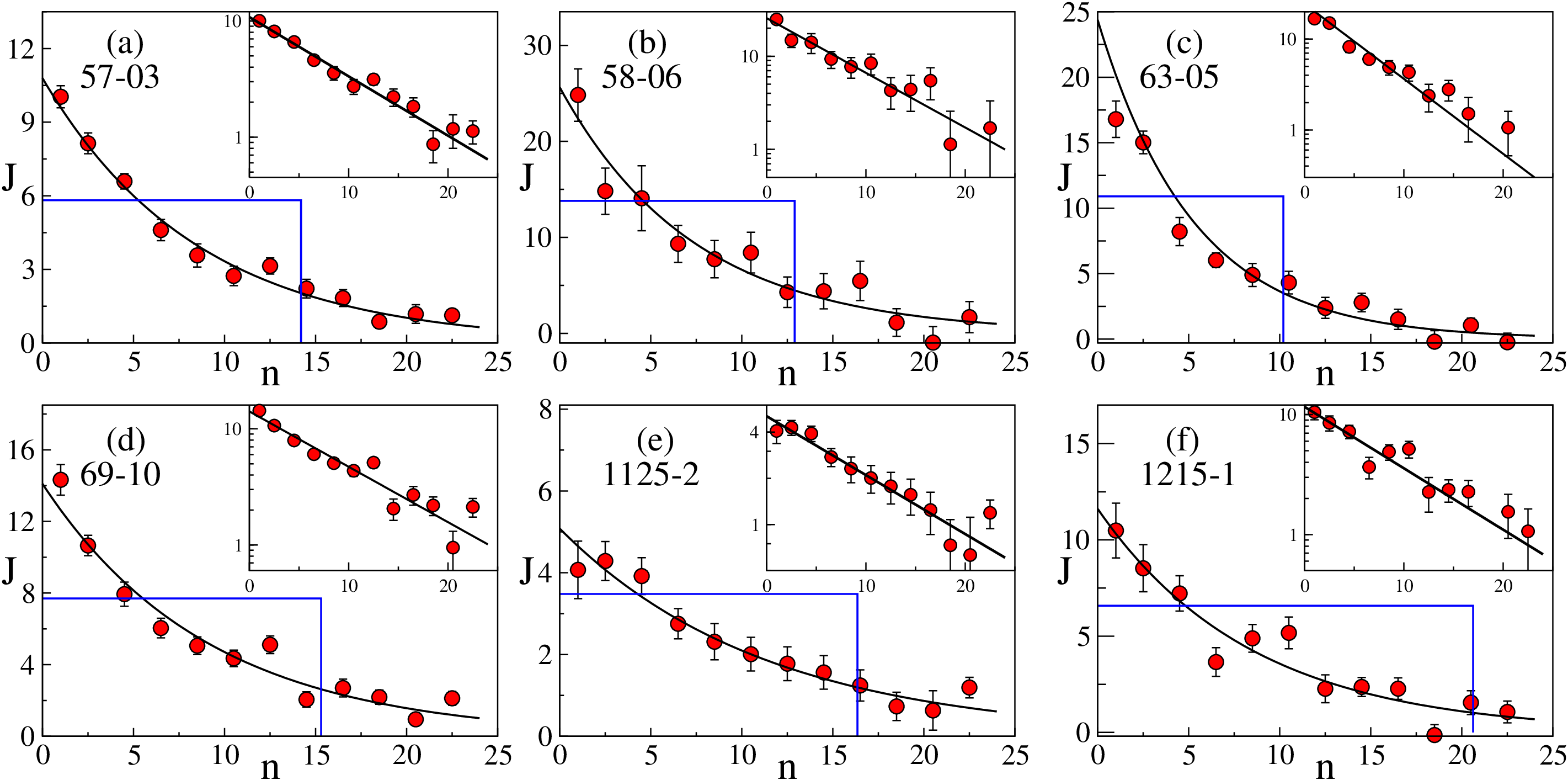}
\caption{The alignment interaction strength $J(n)$ for six events different form those in the main text : event 57-03 (a),
event 58-06 (b), event 63-05 (c), event 69-10 (d), event 20111125-2 (e), and event 20111215-1 (f) (See the Table. \ref{table} for the details of the events). Circular symbols (red) are the result of the ME method, while black solid lines are the exponential fit to the data. Step functions (blue) are the interaction strengths computed using the ME method described in \cite{Bialek12}. Insets are semi-log plots of $J(n)$ for the respective events. All of them show reasonably clear exponential decays.}
\label{fig:morejn}
\end{figure*}

In \fref{fig:fig1}~a,c we can compare the behavior of $C(n)$ (the input) with the inferred effective interaction  strength, $J(n)$ (the output). What we find is that, in contrast with the correlation, the interaction $J(n)$ is very much {\it short-ranged}. The difference between $C(n)$ and $J(n)$ is quite striking (\fref{fig:fig1}, left). We find that  $J(n)$ decays exponentially with the topological distance (\fref{fig:fig1}~b,d),
\bea
J(n) = J_0 \ e^{-n/n_c} \ ,
\label{eqn:newJ}
\eea
where the decay constant $n_c$ provides a measure of the interaction range. The mean value of $n_c$ over all $22$ analyzed flocks is, 
\be
n_c=8.0 \pm 0.5 \quad \mathrm{(std \ error)} \ ,
\ee
to be compared with the estimate $n_c = 6.5 \pm 0.9$ (std error) given in \cite{Ballerini08a} using spatial structure as a proxy of the interaction. Plots of $J(n)$ for several other analyzed events are displayed in Fig.\ref{fig:morejn}, while the values of $n_c$ for all events can be found in Table \ref{table}. In all cases the interaction decays exponentially and the interaction range $n_c$ is much smaller than (and not dependent on) the system's size $N$ (see Table \ref{table}). In terms of metric distances, in all cases these ranges correspond to distances  much smaller than the extension of the flock (and well below its shorter dimension) - see Table \ref{table}. We therefore find  that the effective alignment interaction in starling  flocks is short-ranged \cite{footnote2}.

We notice another interesting aspect of Fig.\ref{fig:morejn}: these plots all have the same scale on the abscissa, meaning that in all flocks the interaction decays
over a similar range of the topological distance $n$. But this flocks have significantly different densities, therefore if we wanted to plot $J$ as a function of  the physical, metric distance $r$ we would need widely different scales. This is yet another demonstration of the previously discovered fact \cite{Ballerini08a, Bialek12} that interaction in bird flocks is based on topological, rather than metric distance \cite{footnote2}.

\subsection{Comparison with the step-interaction case}
\label{sec:onestep}

As mentioned in the Introduction, a first, simpler, maximum entropy computation on bird flocks was performed by some of us in \cite{Bialek12}.  Let us now compare the present  approach with the one of \cite{Bialek12}, and discuss the present step forward compared to previous results.

The experimental input used in \cite{Bialek12} was also related to velocity correlations. However, it was not the correlation as a function of distance ${\hat C}(n)$, as in the present paper. Rather, we considered a single scalar, $\hat{C}_\mathrm{int}$, describing the {\it average} degree of correlation between a bird and its interacting neighbors
\bea
\hat{C}_\mathrm{int} = \frac{1}{N }\sum_i\frac{1}{n_c}\sum^{n_c}_{j \in i} \vec{s}_i \cdot \vec{s}_j,
\label{gufo}
\eea
where the sum is carried out over the first $n_c$ neighbours of $i$.  We can recast this quantity in the language of the present paper by noting that, 
\bea
\hat{C}_\mathrm{int}= \frac{\sum_{i,j=1}^N \vec{s}_i\cdot\vec{s}_j \, \Theta(k_{ij}-n_c)}{\sum_{i,j=1}^N \Theta(k_{ij}-n_c)},
\eea
where $\Theta(x)$ is the Heaviside step function. By using the  formalism that we developed above, 
it is easy to see  that this construction is equivalent to assume that the interaction function has a step-like behavior, being  constant up to neighbour $n_c$ and zero beyond that, namely,
\bea
J(n) = J_0 \ \Theta(n_c-n). 
\label{eqn:oldJ}
\eea
In \cite{Bialek12} $J_0$ - the (average) strength of the interaction - naturally appeared  as the Lagrange multiplier associated to $C_\mathrm{int}$, so that the entropy was maximized w.r.t. it. On the other hand, $n_c$ - the width of the step-like interaction -   was {\it not} the Lagrange multiplier of any given observable,  so it remained in the likelihood even after maximization w.r.t. $J_0$ and was determined through a maximum likelihood principle.  This means that the calculation of \cite{Bialek12} in fact assumed some parameter-dependent form of the model  (namely the step interaction form (\eref{eqn:oldJ})), and did not exclusively rely on entropy maximization.

How does the calculation of \cite{Bialek12} compare to the one we developed above?
First of all, we see from \fref{fig:morejn} that the old step-interaction is always compatible with the new exponential interaction, 
so there is a nice consistency between the two cases.
For a more quantitative comparison, let us  call $J_0^{\mathrm{step}}$ and $n^{\mathrm{step}}_c$ 
the strength and the range of the interaction of the step model of \cite{Bialek12}, 
and $J_0^{\mathrm{exp}}$ and $n^{\mathrm{exp}}_c$ the parameters of the exponential fit of the $J(n)$ that we calculated in the present work (main paper, Eq.~7). It is reasonable to expect two things:
\begin{enumerate}
\item the total interaction strength, that is $\sum _n J(n)$, should be the same in the two cases.
From this condition we get $n_c^{\mathrm{exp}}J_0^{\mathrm{exp}}=n_c^{\mathrm{step}} J_0^{\mathrm{step}}$;

\item if we interpret $w(n) = J(n) / \sum_m J(m)$ as the (normalized) weight of the $n^{th}$ neighbour, 
the average of $n$, i.e. $\sum_n w(n) n$ should be the same in the two cases.
\end{enumerate}
These two conditions give, 
\bea
n_c^{\mathrm{exp}} &=& n_c^{\mathrm{step}}/2    \ ,
\nonumber
 \\
J_0^{\mathrm{exp}} &=& 2J_0^{\mathrm{step}} \ .
\label{zombie}
\eea 

\begin{figure}[t]
\centering
\includegraphics[scale=0.35]{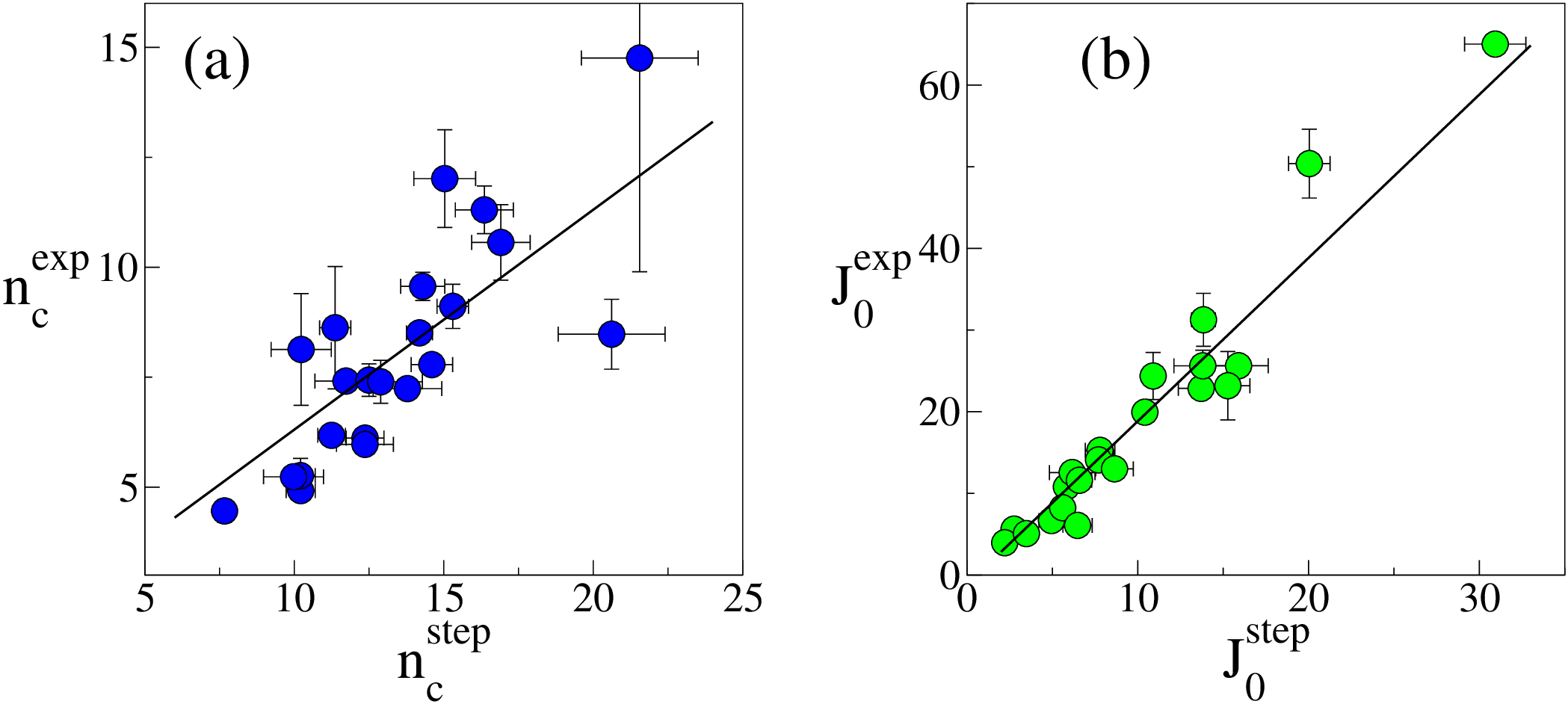}
\caption{
(a) Interaction range $n_c$ and (b) interaction strength $J_0$: Comparison of the step model vs the present work; the full lines are the predictions of (\eref{zombie}).
}
\label{fig:step}
\end{figure}

The data in \fref{fig:step}  indicate that  these two  relations are indeed satisfied. Notice that the fact that the step-like parameter $n_c^{\mathrm{step}}$ is twice as large as the exponential decay range  can (at least partially) explain the discrepancy between $n_c^{\mathrm{step}}$ and the previous estimate of the interaction range given in  \cite{Ballerini08a}.

\subsection{Longitudinal vs transverse interaction}
\label{sec:angular}

In natural flocks the distribution of neighbors around a given individual was found to be anisotropic \cite{Ballerini08a}. This suggests that there might be a certain degree of anisotropy in the interactions between birds. We can use the ME approach to investigate this question. We know that the more detailed the experimental input we use, the more detailed the corresponding ME model will be. In the previous section we discussed how increasing the amount of experimental information can lead to an increase in knowledge about the interactions: using only ${\hat C}_{int}$ \cite{Bialek12} allows inferring an effective interaction range and strength, while using the correlation function ${\hat C}(n)$ allows inferring the full dependence of  interaction on distance. In the same way, to probe the angular dependence of interactions we now consider correlation functions, which not only depend on distance, but also the on angle with respect to the direction of motion.

Given a bird, $i$, we partition the space around it into two sectors, the {\it longitudinal} one and the {\it transverse} one: consider a neighbor $j$ of $i$, and let $\theta_{ij}$ be the angle formed by $\vec r_{ij}$ (the vector joining $i$ to $j$) and the flock's direction of motion $\vec{V}$; then $j$ is in the longitudinal sector of $i$ if $|\cos(\theta_{ij})|>1/2$; otherwise it falls into the transverse sector (this relationship is symmetric). Notice that with this definition the two sectors have the same $3d$ volume.
We then define the longitudinal (L) and transverse (T) correlation functions, which are simply the average correlations in their relative sectors, 
\bea
\hat C^{L,T}(n) = \frac{\sum_{i,j} \vec{s}_i\cdot\vec{s}_j \delta(k_{ij}-n) 
\Theta(\pm|\cos(\theta_{ij})|\mp 1/2)}{\sum_{i,j} \delta(k_{ij}-n) 
\Theta(\pm|\cos(\theta_{ij})|\mp1/2)} \,
\label{eq:correan}
\nonumber
\eea
where $\Theta(x)$ is the Heaviside step function. Interestingly, when computing these correlations on flocks data, we find that the transverse correlation is systematically larger than its longitudinal counterpart at small topological distances.
Starting from these new observables, one can apply the maximum entropy method as explained in the previous sections and get a ME distribution  with effective Hamiltonian;
\bea
H &=& - N \sum_n \left[ p^L(n) J^L(n) \CorrNL + p^T(n) J^T(n) \CorrNT \right], \,
\label{eq:ham-anys}
\eea
Here, the Lagrange multipliers $J^L(n)$ and $J^T(n)$ represent the alignment interaction strengths of a bird with its $n$-th neighbour in the longitudinal and transverse direction respectively. The  $p^{L,T}$ are the fraction of neighbors that lie in longitudinal and transversal sector and are defined as $p^{L,T}(n) = (1/N) \sum_{i,j} \delta(k_{ij}-n) \Theta(\pm|\cos(\theta_{ij})|\mp1/2).$ 
These quantities of course satisfy the relation $p^L(n)+p^T(n)=1$. We note that, despite the constrains:
\bea
\CorrN &=& p^L(n) \CorrNL + p^T(n) \CorrNT,
\label{eqn:cnstrCLTn}
\eea
the link between $J(n)$ and $J^{L,T}(n)$ is not trivial. In this case we match at the same time $\CorrNL$ and $\CorrNT$, in the isotropic case $\CorrN$ only. But $\CorrNL$ and $\CorrNT$ can be very different from each other while maintaining $\CorrN$ same (\eref{eqn:cnstrCLTn}). This means that we can have many combinations of $J^{L,T}(n)$ consistent with the same $J(n)$. Special cases occur only when the correlation functions of the two sectors are the same $\CorrNL = \CorrNT$ (in this case $J^L(n)=J^T(n)=J(n)$) or when there are no neighbors in one of sectors (if $p^T(n)=1$ then $J^T(n)=J(n)$ and $J^L(n)$ is indeterminate, and viceversa -- as it should be).

To find the $J^{L,T}(n)$ consistent with experimental data we proceed along the lines explained in the previous sections.
Also in this case the Hamiltonian (\ref{eq:ham-anys}) can be recast in an Heisenberg-like form, which allows computing analytically the partition function to get an explicit expression of the log-likelihood in terms of $J^L(n)$ and $J^T(n)$ (see Appendix \ref{app:coarse}). The transverse and longitudinal interaction functions can then be retrieved by maximizing the log-likelihood. 

\begin{figure}
\centering
\includegraphics[width=0.4\textwidth]{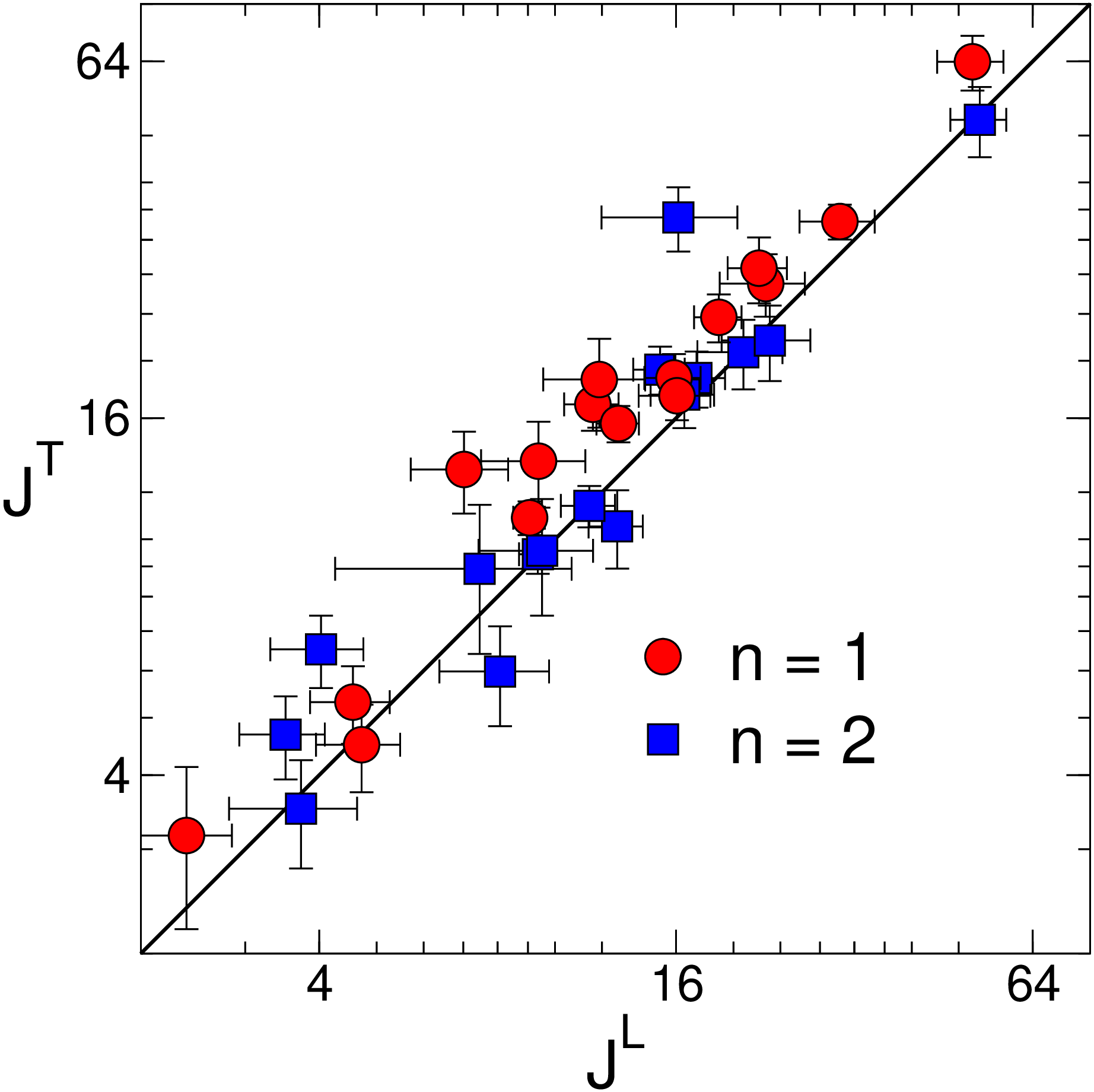}
\caption{Log-log plot of $J^L$ vs $J^T$ for $n=1$, and $n=2$ for all the analyzed 
flocks. The full line is the identity. The computation of both correlations and inferred interactions in the anisotropic case requires a larger statistics, because only half of birds pairs are on average used to get $J^{L,T}(n)$ and ${\hat C}^{L,T}(n)$. For this reason, events where the size is too small or that are too short in time are too noisy and have been not included in the analysis (events 77-07, 72-02, 1214-4-1,1214-4-2, 58-07 - see Table \ref{table})}
\label{fig:fig3}
\end{figure}

The result is shown in \fref{fig:fig3}, where we plot the values of $J^T(n)$ and $J^L(n)$ for $n=1$ and $n=2$ for all the analyzed flocking events. The interaction between nearest neighbors in the transverse direction is detectably stronger than that in the longitudinal direction. 
On an average, $J^T(n=1)$ is $20 \%$ larger than $J^L(n=1)$. More precisely, $J^T(1)$ and $J^L(1)$ are linearly correlated with  Pearson correlation coefficient $\rho > 0.99$, the best linear fit giving $J^T(1) = (1.208 \pm 0.058) J^L(1)$  (statistical confidence p-value $< 1.0e-6$). 
This anisotropic character of the interaction is very short-ranged, though, as it already disappears by the second nearest neighbor, $J^T(n=2)\simeq J^L(n=2)$ (\fref{fig:fig3}).  

The anisotropy that we find is not strong, but it is interesting. Let us discuss more in details its possible origins and consequences. 

First, remember that we are studying the {\it alignment} interaction, hence our result tells us that a bird is more keen to align its direction of motion with the neighbor on the side, rather than with that directly in the front. One may speculate that this is due to the fact that misalignment with a side neighbor has more severe consequences (in terms of collision) than that with someone along the direction of motion. On the other hand, for what concerns {\it speed} control one would expect the opposite: a stronger interaction in the longitudinal direction would be more useful to avoid bumping into each other. Some recent progress has been made in working out the speed interaction in flocks \cite{Bialek14}; it would therefore be interesting to extend the present calculation to the case of speed.

Even though the anisotropy concerns the directional degrees of freedom (the velocities) it can have an impact on the spatial arrangement of neighbors. One can argue that individuals who better coordinate flight directions tend to keep the same mutual distance, and consistently maintain their neighborhood relationship. In this respect, our result is consistent with the finding  of \cite{Ballerini08a}, where it was found that the closest neighbors of a bird are more easily found in the transverse than in the longitudinal direction (i.e. the nearest neighbors are typically on the side rather than in the direction of motion). Understanding how interactions between flight directions are related to the spatial structure of the group is a complex problem. There could be positional attraction-repulsion forces between birds, which we did not consider in our ME analysis, and that can influence the spatial arrangement of individuals (see \cite{Charlotte12,franck} for a discussion in numerical models). Recent analysis \cite{Castellana15} however suggest that in systems with  topological interactions velocity alignment  has an important role in the structure, which is why our result can help to understand this issue.

Finally, a word of caution is required. There is an important anisotropy present in polar active systems, which is a consequence of symmetry breaking and dynamics and is not due to anisotropic microscopic interactions. Flocks are polarized groups, as such velocity fluctuations orthogonal to the global velocity are much stronger than longitudinal fluctuations, due to the presence of soft modes (see Appendix \ref{app:spinwave}).
In self-propelled systems this causes anisotropic diffusion exponents and a non trivial scaling of correlations in the large scale hydrodynamic regime \cite{Toner98}. Natural flocks exhibit their collective behavior on much shorter scales - both in terms of sizes and time \cite{Attanasi13a}. Still, it might be that the anisotropy captured by the ME model in part describes the effect of the microscopic anisotropic diffusion on the scale of the experimental observations.

\section{Dependence on the number of input  variables}
\label{sec:nmax}

When we introduced the ME approach in Sec.\ref{sec:maxent} we briefly discussed the role of the number of  input experimental observables. The more observables we consider, the more detailed the corresponding ME model. Indeed, the number of parameters that we infer through the method is equal to the number of experimental observables that we use to constrain the entropy maximization. In principle, increasing the amount of experimental input should lead to a more complete knowledge of the effective interactions in the system (see e.g. Sec.\ref{sec:onestep}, \ref{sec:angular}). This is not however always true. There are cases where the relevant information is captured by a small number of observables, and one just needs these few observables to get a very effective description of the system. Our case is precisely of this kind, and offers a very nice example where the predictive power of a ME model can be clearly quantified.

In our work we use the correlation function $\CorrN$ as reference observable.
The topological distance $n$ between two birds can go up to $n_\mathrm{max}=N$. This means that the correlation function (\ref{eqn:cn}) is a set $N$ numbers, so that we should in principle use $N$ Lagrange multipliers, $J(n)$ with $n=1,\dots,N$, to maximize the entropy. The very long-range nature of $C(n)$ seems to suggest that there is indeed information to be exploited in this whole function, up to the maximum possible values of the topological distance. In fact, the situation is very different.

\begin{figure}
\centering
\includegraphics[width=0.7\textwidth]{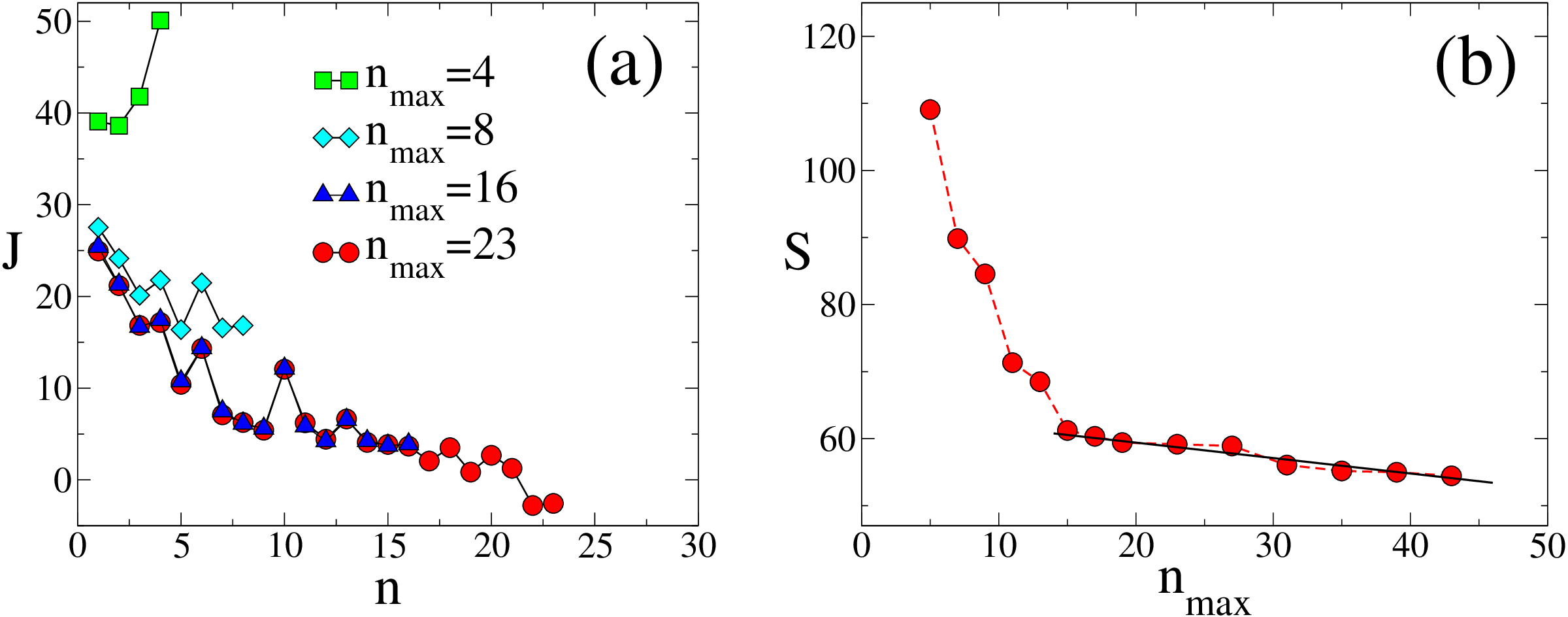}
\caption{(a) The interaction strength $J(n)$ is plotted for different $n_{max}$ (event 21-06). 
With increasing $n_\mathrm{max}$, the form of $J(n)$ saturates. 
(b) Entropy $S$ vs $n_\mathrm{max}$ for the same event. In the large $n$ regime the entropy decays very weakly; in this regime we are merely fitting the noise.}
\label{fig:fig2}
\end{figure}

What we find is that if we use values of $\hat{C}(n)$ up to a maximum distance $n_\mathrm{max}$, the inferred interaction stabilizes for $n_\mathrm{max} \ll N$. To see this we maximized the entropy for different numbers of Lagrange multipliers, that is we calculated $J(n)$, with $n=1,\dots ,n_\mathrm{max}$, for different values of $n_\mathrm{max}$ (\fref{fig:fig2}a). What we see is that for very small $n_\mathrm{max}$ the function $J(n)$ is unstable, so that the whole interaction changes drastically when increasing $n_\mathrm{max}$. However, when $n_\mathrm{max}$ becomes large enough the full interaction $J(n)$ stops depending on $n_\mathrm{max}$ and the only effect of feeding more correlations and adding new parameters to the calculation is to obtain negligible and noisy couplings. This means that beyond a certain distance, the ME calculation simply refuses to switch on any more couplings, even though the long-range correlation function that we feed as an input still seems ripe of information at that distance. 
This is an indication that the ME method works with remarkable economy.

The role of $n_\mathrm{max}$ can be understood also at the level of the entropy. The value of the entropy as a function of $n_\mathrm{max}$ after maximization  tells us how much information we gain \cite{Shannon48} by adding more experimental data ($\hat C(n)$) and by inferring more parameters ($J(n)$). We can see from \fref{fig:fig2}b that the entropy decays very fast up to a certain $n_{max}\simeq15$, and then the decay becomes slower and linear. This means that we are gaining real information for $n_\mathrm{max}\leq15$, but after that we are simply fitting the noise and there is no more useful information to be gained by increasing $n_\mathrm{max}$. For infinitely large $N$, that is for infinitely accurate experimental averages, we expect the large $n$ weak decrease of the entropy to become a real plateau, signifying that there is really nothing to gain (not even in terms of noise fitting) by adding more parameters than those really required by the short-range interaction.

\begin{figure}[b]
\centering
\includegraphics[scale=.3]{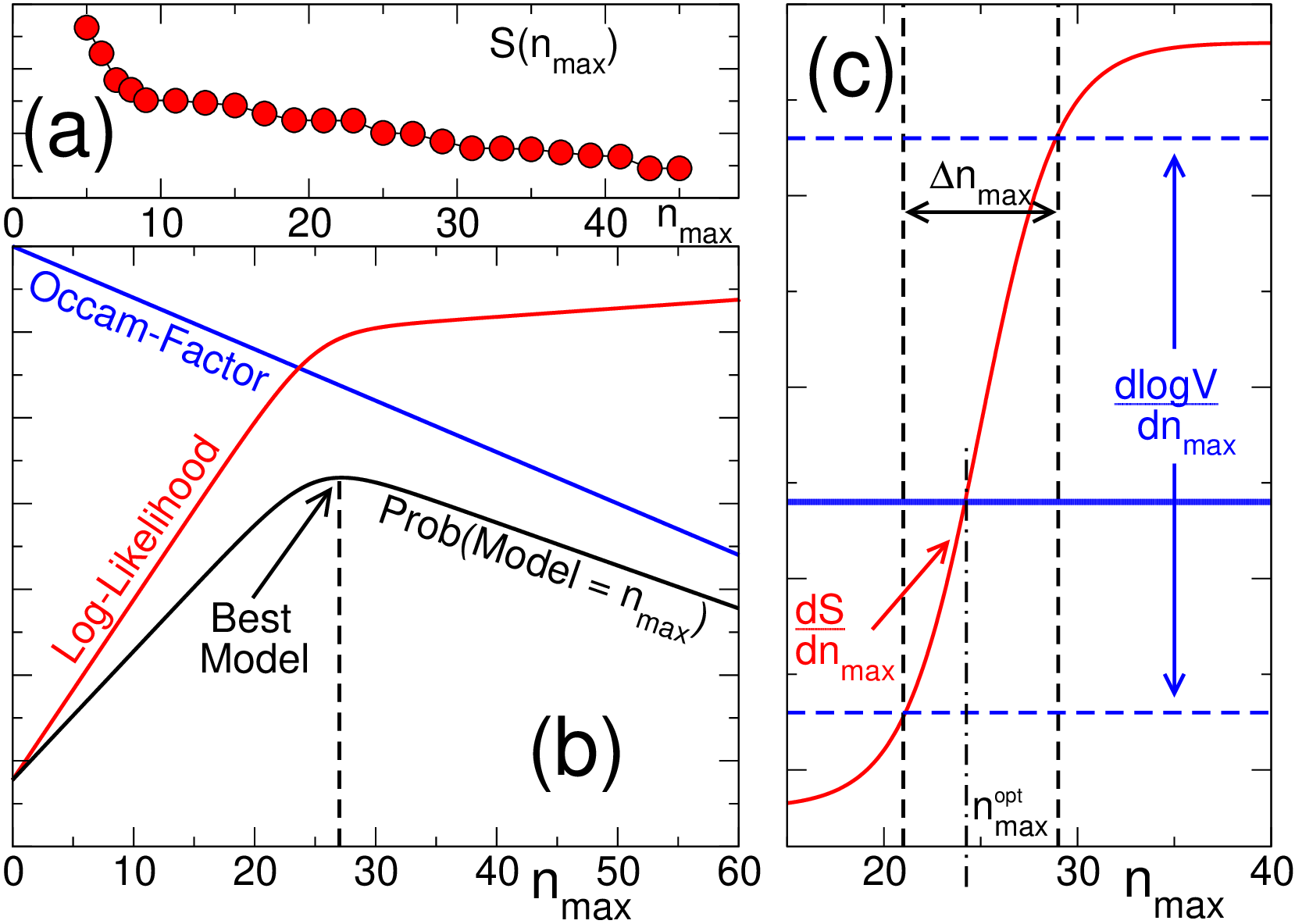}
\caption{
(a): Entropy as a functions of number of parameters for one real event.
(b): Contributions of log-likelihood and Occam factor to probability of model $P(n_\mathrm{max}|D)$.
(c): Red line is derivative of entropy and
   each horizontal blue lines represents derivative of logarithm of Occam factor 
   for different values of prior.
}
\label{fig:entropy}
\end{figure}

To better understand the role of the entropy and fully appreciate the meaning of the change of behavior displayed in \fref{fig:fig2} we need a Bayesian analysis. If we want to infer a good model, a model that tells us something about the behavior of the system, our purpose is not simply to fit well the data. Rather, our goal is to find  the {\it minimal} set of parameters able to reproduce  the experimental data. Within a Bayesian framework this goal can be mathematically formalized. Let us call $P(n_\mathrm{max}|D)$ the probability of a model with $n_\mathrm{max}$ parameters, given a certain dataset $D$. It can be shown that \cite{Mackay}
\be
P(n_\mathrm{max}|D)=P(D | n_\mathrm{max}) V(n_\mathrm{max}) 
\sim e^{-S(n_\mathrm{max})} 
e^{-\alpha n_\mathrm{max}} \ .
\ee
The first term in the r.h.s. is the maximized likelihood, i.e. the probability of getting the data with a model with $n_{max}$ parameters, and is given by the exponential of the ME entropy (\eref{eqn:loglikelihoodfull0}).
The second term, which is called Occam factor, $V(n_\mathrm{max})$, is equal to the 
ratio between the posterior accessible volume in the space of parameters and the prior accessible volume  \cite{Mackay}.
Typically, the Occam factor decays exponentially with the number of 
parameters, $V(n_\mathrm{max})\propto e^{-\alpha n_\mathrm{max}}$.

Hence, in general when we increase the number of parameter $n_\mathrm{max}$ of the model we have a trade-off: 
on one hand it improves the fit, hence it increases the likelihood;
on the other hand, it decreases the Occam factor.
Because of this trade-off, when the number of parameters increases beyond a certain value,
the suppressing contribution of the Occam factor compensates the decay of the entropy, and therefore the growth of the likelihood.
For this reason $P(n_\mathrm{max}|D)$ reaches a maximum for a finite value of parameters, $n_\mathrm{max}=n_\mathrm{max}^{opt}$ \fref{fig:entropy}(b).

Unfortunately, the Occam factor depends on the {\it prior} probability of the parameters, which is always an obscure thing. For this reason 
the position of this maximum is not clearly defined. However, it is possible to show that this fact produces only a small ambiguity in the location of $n_\mathrm{max}^{opt}$.
First of all, the contribution of the Occam factor to $P(n_\mathrm{max}|D)$ depends only logarithmically on the prior probability: 
major changes in the prior leads to a small change in Occam factor.
Moreover, when we increase $n_\mathrm{max}$ after we reach the optimal value, the slope of entropy changes suddenly:
in fact, we move from the regime $n_\mathrm{max}<n_\mathrm{max}^{opt}$, where adding each new observable implies a considerable increase of information,
to the regime $n_\mathrm{max}>n_\mathrm{max}^{opt}$, where instead adding  new observables only marginally increase the total information.
$n_\mathrm{max}^{opt}$ is determined by the condition ${\partial S}/{\partial n_\mathrm{max}} = {\partial \log{V(n_\mathrm{max})}}/{\partial n_\mathrm{max}}$, which means that the solution is the crossing point  between red and blue line in \fref{fig:entropy}(c).
Varying the prior, the blue line moves up and down 
and this moves $n_\mathrm{max}^{opt}$ by an amount $\Delta n_\mathrm{max}$.
As we can see from the figure, the faster the change of slope of entropy, the smaller the range $\Delta n_\mathrm{max}$.
Then, typically, big changes in the prior leads to little change in $n_\mathrm{max}^{opt}$.


\section{Conclusions}
\label{sec:conclusions}
In this paper, using the ME approach, we have provided rather direct evidence that the effective alignment interaction between starlings within real flocks is short-range. This result is interesting for two reasons.

First, from a biological perspective, we believe this is the first time that short-range interaction is found without being an  a priori 
ingredient of the model used to fit the data. In general, it is difficult to formulate a model where a qualitative crossover from short to long-range interaction occurs by tuning a parameter. Hence, what is normally done is that a certain, fixed, functional form is assumed, and its parameters fitted.  Here, on the other hand, we assumed no  a priori functional form of the interaction, so that the final result is completely ruled by the experimental data. We believe that short-range interaction (at least in starling flocks) can now be considered as a rather well established fact.

Secondly, our result is relevant for the maximum entropy method itself, which is increasingly used in biological inference \cite{Schneidman06, Shlens06, Lezon06, Tang08, Weigt09, Halabi09, Mora10,TkacikReview10, Tkacik13,Tkacik14}. A common objection to the ME method is that it is just another kind of model fitting procedure, so that, ultimately, one is prone to obtain as a {\it qualitative} output of the method what one feeds into the method. We believe that what we have found here proves otherwise. The difference between long-range and short-range interaction {\it is} a qualitative one, with deep consequences on the physics of the system. Yet we have seen that long-range correlation is turned into short-range interaction by the ME method, with the entropy pointing out what is the minimal number of parameters that need to be switched on, given the data. This suggests  that the maximum entropy method manages to extract information from a data set with minimal bias.

We thank William Bialek for many discussions and suggestions. This work was supported by grants IIT--Seed Artswarm, ERC--StG n.257126 and US-AFOSR - FA95501010250 (through the University of Maryland).

\appendix

\section{Partition function in spin-wave approximation:}
\label{app:spinwave}
In order to calculate the log-likelihood we need to compute the partition function $Z$ (\eref{eqn:loglikelihoodfull0}).
In general, the exact analytical calculation of the partition functions is very hard. 
In the case of flock, however, this can be done thanks to the spin-wave approximation \cite{Bialek12}. 
The idea is that, because flocks are very ordered, with magnetization (i.e. polarization) close to $1$, 
we can expand in the small fluctuation around the mean direction of motion.
The partition function can be written as,
\bea
Z &=& \int D\vec{s} \lx[ \prod_i \; \delta(|\si|-1) \rx] \exp \lx[\sumij J_{ij} \si \cdot \sj \rx],
\label{eqn:partition}
\eea  
where $D\vec{s} = \prod_i d\si$, and the $\delta-$function 
is enforcing the constraint that each spin has unit length.   
We define global order parameter, $\vec{V} = \sumi \si / N = \Phi \,\hat{n}$, 
where $\hat{n}$ is unit vector and $\Phi=|\vec{V}|$ is the  polarization of the flock.
Each spin $\si$ can be rewritten in terms of the global orientation 
direction $\hat{n}$ and a perpendicular component to $\hat{n}$, that is $\si = s_i^{L}\hat{n}+\piv{i}$. By construction, 
they satisfies the following relations:
\bea
\piv{i} \cdot \hat{n} = 0\ ,
\qquad
\frac{1}{N}\sumi {s_i^L} = \Phi \ ,
\qquad
\sumi \piv{i} = 0 \ .
\eea
The $\piv{i}$ are the small fluctuations that  the spin-wave approximation uses to expand the partition function.
The partition function can be rewritten as,
\bea
Z &=& \int Ds^L D\vec{\pi} \lx[ \prod_i \delta\lx(\sqrt{(s_i^L)^2 + |\piv{i}|^2-1}\rx) \rx] 
\delta\lx(\sumi \piv{i} \rx) \exp \lx[\sumij J_{ij} (s_i^L s_j^L + \piv{i} \cdot \piv{j} ) \rx],
\eea 
where $Ds^L = \prod_i ds_i^L$ and  $D\vec{\pi} = \prod_i d\piv{i}$.
The delta functions are taking care of the constraint 
on the length of each spin and of the global constraint on the $\piv{i}$. 
For strongly ordered flock, $\Phi \simeq 1$ and $|\piv{i}| \ll 1$. 
Then, at second order, $s_i^L \simeq 1- |\piv{i}|^2/2$. 
Performing the integral over $s^L$, the partition function becomes,
\bea
Z &=& \int D\vec{\pi} \lx[ \prod_i \frac{1}{\sqrt{1-|\piv{i}|^2}} \rx] \delta\lx(\sumi \piv{i}\rx) 
\exp \lx[ -\sumij A_{ij}\piv{i} \cdot \piv{j} + \sumij J_{ij}\rx],
\eea
where
\bea
A_{ij} &=&  \delta_{ij} \lx(\sum_k J_{ik}\rx) -  (1 - \delta_{ij}) J_{ij}.
\label{eqn:aij}
\eea
For strongly order flock, the product $\prod_i 1/\sqrt{1-|\piv{i}|^2}$ can be neglected 
(we have explicitly checked  that the corrections due to this term are indeed negligible).
Therefore, we can write
\bea
Z &=& \int  D\vec{\pi} \; \delta \lx( \sumi \piv{i} \rx) \; 
 \exp \lx[ - \sumij A_{ij} \piv{i} \cdot \piv{j} + \sumij  J_{ij} \rx].
\eea
Because $J_{ij}=J_{ji}$ (and then $A_{ij}=A_{ji}$) we benefit from the spectral theorem for symmetric matrices.
The matrix $A_{ij}$ is diagonalizable, its eigenvalues are real and its eigenvectors form an orthonormal basis.
Moreover, the condition $\sumj A_{ij} = 0$ means that the matrix $A_{ij}$ is a positive semidefinite matrix,
in particular, the smallest eigenvalue is $a_1=0$, and all other eigenvalues are positive. 
Let  $a_k$ be the eigenvalue of the eigenvector ${\bf w}_k$.
The eigenvector ${\bf w}_k$ satisfies the usual relation,
\bea
\sumj A_{ij} w_j^k = a_k w_i^k.
\eea 
It can be easily seen that the eigenvector ${\bf w^1}$ corresponding to 
$a_1$ is constant and it is given by $(1/\sqrt{N},1/\sqrt{N}, .., 1/\sqrt{N})$.
We can rewrite the integral in orthonormal basis defined by ${\bf w}^1, {\bf w}^2,...,{\bf w}^N$ :
\bea
Z &=& \int D\vec{\pip} \; \delta(\pipv{1})
\exp \lx[ -  \sum_{k=1}^{N} a_k |\pipv{k}|^2  + \sumij J_{ij} \rx],
\eea  
where $\pipv{k} = \sumi w_i^k \piv{i}$. 
From this form it is clear that the role of the $\delta$-function 
over the $\pipv{1}$ is exactly to eliminate the zero mode from the integral.
Performing the Gaussian integral in two dimensions, we obtain,  
\bea
\log{Z} &=& -\sum_{k>1} \log{a_k} + \sumij J_{ij} , 
\label{eqn:logZ}
\eea
where the irrelevant constant terms have been neglected. 

\section{Coarse-graining}
\label{app:coarse}

The experimental observable we consider in flocks is the correlation function $\CorrN$.
In principle, this correlation function can be computed for each value of the topological distance $n$ but, 
as discussed in the paper, it is safe to consider only the observables $\CorrN$ up to $n=\nmax \ll N$.
Furthermore, in order to decrease the number of parameters and speed up the numerical task,
we can consider a ``coarse graining'', that is a binning of $n$ with generic increment $\Delta n \ge 1$.
This means that we include in the same observable $\CorrN$ contributions from the distances $n,n + 1,...,n + \Delta n - 1$.
Hence, we have, 
\bea
\CorrN &=& \frac{\sumij \si \cdot \sj \  \hdkijmn}{\sumij \hdkijmn},
\eea
where $\{\si\}$ are unit vectors and $\kij = n$ if $j$ is the $n^{th}$ neighbor of $i$.
$\delta(k - n)$ is a ``modified'' Kronecker's $\delta$ that takes into account the binning of $n$,
\bea
\delta(k - n) &=& \lx\{
\begin{array}{l l}
    1 & \text{if} \quad n \le k <  n + \Delta n,\\
    0 & \text{otherwise}.
\end{array}\rx.
\label{eqn:coarsedelta}
\eea
Note that for $\Delta n=1$ the model reduces to the one described in the main text. 

For each observables $\CorrN$ the associated Lagrange multiplier is denoted by $\lambda_n$.
The maximum entropic Hamiltonian consistent with these observables is
\bea
H &=& \sumpn \lambda_n \CorrN,
\label{eqn:HJn}
\eea
where the symbol $\sum^{'}$ means that the sum stops at $\nmax$ and that we sum only over the ``bins'' of the coarse graining, that is $n=1,1+\Delta n, 1+2\Delta n, ...,\nmax$.
Physically, $\lambda_n$ is the {\it total} interaction strength for bin $n$ for a flock (extensive), whereas
the interaction strength $J(n)$ defined in main text is the strength for an {\it individual pair} within bin $n$ (intensive) and hence $J(n) = -\lambda_n / (N \Delta n)$.  

Using coarse-grained variables does not change formally the computation, as outlined in  the main text.
  Indeed if we define $\hat{J}(\kij)$ such that
\bea
\hat{J}(\kij) &=& \sumpn J(n) \hdkijmn,
\eea
(note that with $\Delta n = 1$, $\hat{J}(\kij) \equiv J(\kij)$),
we can easily verify that \eref{eqn:HJn} reads as a classical Heisenberg model 
\bea
H(\{\si\}) \; & = &\; - \sumij \hat{J}(\kij) \; \si \cdot \sj \equiv \;-\StdHmlt. \; \;
\label{eqn:heisenberg}
\eea
where the symmetrized interactions matrix $J_{ij}$ are given by
\bea
J_{ij} \; &\equiv& \;  \frac{1}{2}[\hat{J}(\kij) + \hat{J}(\kji)] =\;  \frac{1}{2} \sumpn J(n) [\hdkijmn + \hdkjimn],
\label{eqn:jijfull}
\eea
The partition function can be computed using the spin-wave expansion, as described in Appendix \ref{app:spinwave}, where, now, the eigenvalues $a_k$ refer to the matrix $J_{ij}$ in Eq.~(\ref{eqn:jijfull}). The log-likelihhood takes the form
\be
\log{\mathcal{L}} = -\log{Z} -\lambda_n \CorrNexp = -\log{Z} + N \Delta n \sumpn J(n) \CorrNexp 
\ee

The coarse graining procedure can be easily generalized to the anisotropic case. We consider the transverse and longitudinal coarse grained correlations (see Eqs.~(\ref{eq:correan})) 
\bea
\CorrNL &=& \frac{\sumij \si \cdot \sj  \hdkijmn \ThetaL}{\sumij \hdkijmn \ThetaL},\\
\CorrNT &=& \frac{\sumij \si \cdot \sj  \hdkijmn \ThetaT}{\sumij \hdkijmn \ThetaT},
\eea
where, we remind, $\theta_{ij}$ is the angle formed by $\vec{r}_{ij}=(\vec{r}_j - \vec{r}_i)$
and the flock's direction of motion $\vec{V}=1/N\sum_i \si$. $\Theta(x)$ is the Heaviside step function and the factor $1/2$ divides the space evenly between the two sectors. The $\delta$-function bears the same meaning as in  (\eref{eqn:coarsedelta}) and identify pairs belonging to the same bin centered around the topological distance $n$ and of width $\Delta n$.

Following the same method as above, when using coarse grained correlations we need to introduce different Lagrange multipliers $\lambda^{L,T}_n$ for each bin (rather than for each discrete value of $n$).  
The ME Hamiltonian then reads
\bea
H &=& \sumpn \lambda_n^L \CorrNL + \lambda_n^T \CorrNT 
\eea 
Also this Hamiltonian can be written as an Heisenberg-like Hamiltonian. The procedure is slightly more complicated than in the isotropic case. 

We first define the fraction of neighbors that lie in longitudinal and transversal sector for each bin around $n$,
\bea
&p^L(n) = \frac{1}{N \Delta n}\sumij \hdkijmn \ThetaL,\\
&p^T(n) = \frac{1}{N \Delta n}\sumij \hdkijmn \ThetaT,\\
\eea
These quantities of course satisfy the relation $p^L(n)+p^T(n)=1$. At this point, we can express the Lagrange multipliers $\lambda_m^{L,T}$ (defined for a given bin)  in terms of the effective longitudinal and transversal interactions $J^{L,T}(n)$  (defined for of each pair of individuals at distance $n$). We have  $J^{L,T}(n)~=~-~\lambda_n^{L,T}/(p^{L,T}(n)N\Delta n)$.
Then, as above, we introduce the pairwise interactions $\hat{J}^{L,T}(\kij)$
\bea
\hat{J}^{L,T}(\kij) &=& \sumpn J^{L,T}(n) \hdkijmn \Theta^{L,T}_{ij},
\eea
where $\Theta^{L,T}_{ij} \equiv \Theta(\pm |\costij| \mp 1)$ and selects pairs that contribute to, respectively, the longitudinal and the transverse sectors.
With these substitutions the Hamiltonian acquires an Heisenberg form, 
\bea
H(\{\si\}) \; &=& \; - \sumij \lx[ \hat{J}^L(\kij) + \hat{J}^T(\kij) \rx] \si \cdot \sj
 \equiv  \;-\StdHmlt
\eea
where  $J_{ij}$ is now the symmetric part of the matrix $\hat{J}^{L}(\kij)+\hat{J}^{T}(\kij)$

The log-likelihhood then becomes
\bea
\log{\mathcal{L}} &=& -\log{Z} -\lambda_n^T \CorrNTexp -\lambda_n^L\CorrNLexp 
\nonumber \\
&=& -\log{Z} +  N \Delta n  \sumpn \left [ p^T(n) J^T(n) \CorrNTexp)  + p^L(n) J^L(n)  \CorrNLexp)\right].\;
\eea

All figures displayed in this paper are obtained using a coarse graining with $\Delta n=2$ for $n=2,\cdots n_\mathrm{max}$, and $\Delta n=1$ for $n=1$. We also used $\Delta n=1$, results are fully consistent with the larger coarse graining, just more noisy.

\section{Computing the derivatives of the Log-Likelihood}
\label{app:der}

Using the analytical expression for partition function \eref{eqn:logZ}, we can write
the expressions of the log-likelihood  (\eref{eqn:loglikelihood}) for the case of full interaction,
\bea
\log{\mathcal{L}} &=& -\log{Z} + N \Delta n \sumpn J(n) \CorrNexp 
= \sum_{k>1} \log{a_k} - N \Delta n  \sumpn J(n) (1 - \CorrNexp). \;
\label{eqn:loglikelihoodfull}
\eea
Similarly, the log-likelihood function for anisotropic case is given by,
\bea
\log{\mathcal{L}} &=& \sum_{k>1} \log{a_k}
-  N \Delta n \sumpn p^T(n) J^T(n) (1 - \CorrNTexp) 
 - N \Delta n \sumpn p^L(n) J^L(n) (1 - \CorrNLexp).\;
\label{eqn:loglikelihoodanis}
\eea
The condition for maximizing the log-likelihood is $\partial{\log{\mathcal{L}}}/\partial{J(n)} = 0$.
Let us consider, for the moment, the isotropic case.
The derivative of the second term of the log-likelihood with respect to $J(n)$ is trivial.
However, differentiating the partition function is far less trivial, 
as the eigenvalues $a_k$ are very complicated functions of the $\{J(n)\}$. 
We will calculate $\partial a_k/\partial J(n)$ by using perturbation theory.
Suppose that we perturb $J(n)$ by some small amount,
$J(n) \rightarrow  J(n) + \epsilon$, where, $\epsilon$ is infinitesimal.
The perturbation makes $A_{ij}$ change into,
\bea
\widetilde{A}_{ij}(\epsilon) &=& A_{ij} + \epsilon \gamma_{ij}(n), 
\eea
where we introduced a symmetric matrix
\bea
\gamma_{ij}(n) &=& \frac{\partial A_{ij}}{\partial J(n)} = 
\sum_{l,m} \frac{\partial A_{ij}}{\partial J_{lm}} \frac{\partial J_{lm}}{\partial J(n)}.
\label{eqn:gamma}
\eea
Due to this small perturbation the eigenvalue $a_k$  and its eigenvector ${\bf w}^k$ change by small amount, 
\bea
\widetilde{a}_k(\epsilon)   &=&  a_k  + \epsilon \xi_k + \mathcal{O}(\epsilon^2), \\
\widetilde{w}_i^k(\epsilon) &=&  w_i^k + \epsilon g_i^k + \mathcal{O}(\epsilon^2).
\eea 
For the $\widetilde{A}(\epsilon)$ matrix we can write,
\bea
\sumj \widetilde{A}_{ij}(\epsilon) \widetilde{w}_j^k(\epsilon) &=& \widetilde{a}_k (\epsilon) \widetilde{w}_i^k (\epsilon).
\eea

\begin{figure}[]
\centering
\includegraphics[scale=.25]{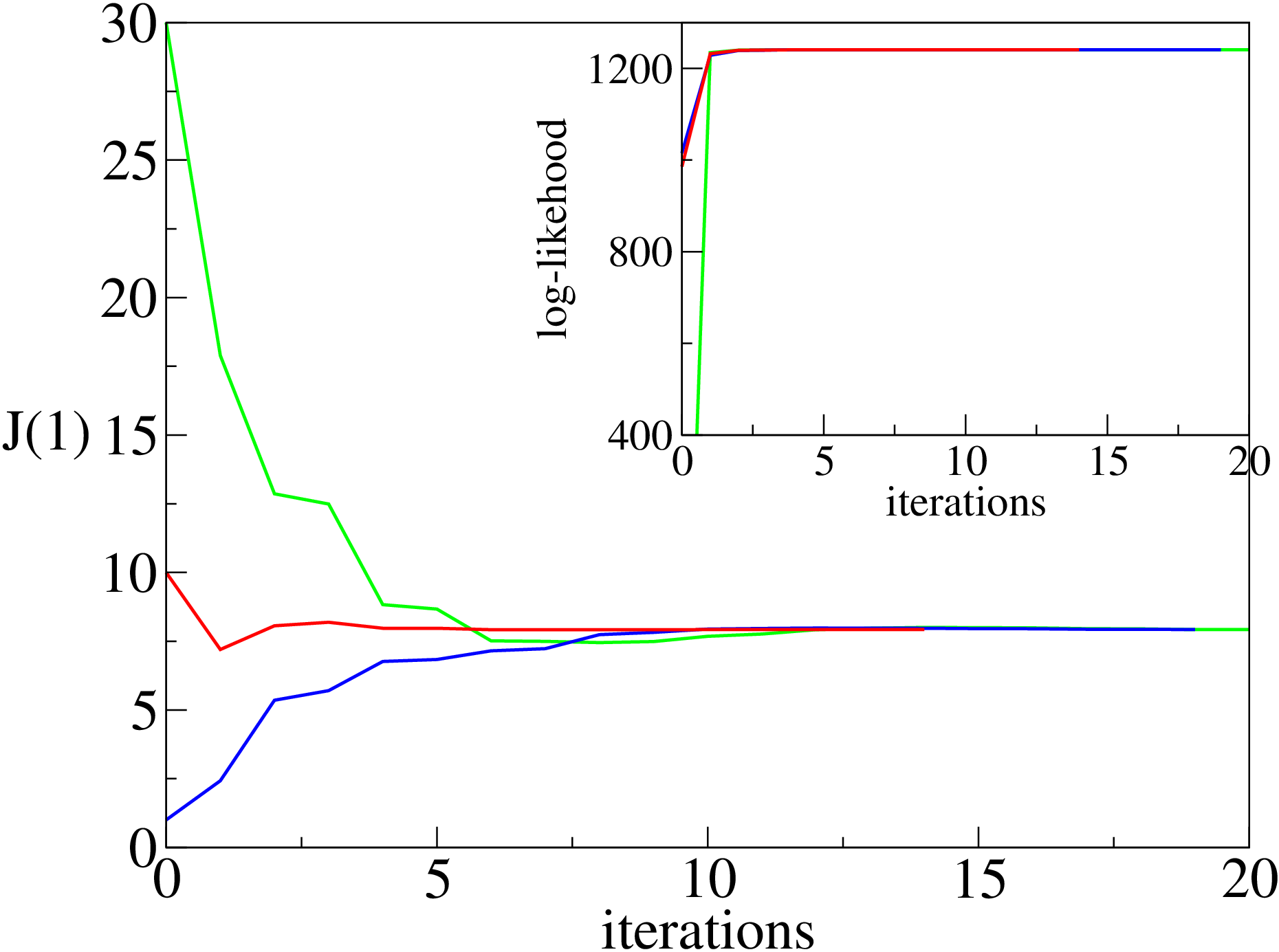}
\caption{Stability and convergence of the numerical method:
three very different initial guesses lead to the same value of $J(1)$
(this is generically true for all other $J(n)$).
Inset: also the log-likelihood reaches the same asymptotic value with different initial conditions.
This implies that it exists a stable and unique global maximum of the log-likelihood.}
\label{fig:iteration}
\end{figure}

Through some algebra it is quite straightforward to show that,
\bea
\widetilde{a}_k(\epsilon) &=& a_k + \epsilon \sum_{ij} \gamma_{ij}(n) w^k_i w^k_j + \mathcal{O}(\epsilon^2).
\eea
Therefore, the derivative of the eigenvalue $a_k$ can be written as 
\bea
\frac{\partial a_k}{\partial J(n)} &=&
\lim_{\epsilon\rightarrow 0} \frac{\widetilde{a}_k(\epsilon) - a_k}{\epsilon} = \sumij \gamma_{ij}(n) w_i^k w_j^k .
\label{eqn:deval}
\eea
To obtain the form of matrix $\gamma_{ij}(n)$ we use first \eref{eqn:aij} and \eref{eqn:jijfull} from which
\bea
\frac{\partial A_{ij}}{\partial J_{lm}}  &=& \delta_{il} \lx(\delta_{ij} - (1 - \delta_{ij}) \delta_{jm}\rx) \\
\nonumber
\frac{\partial J_{lm}}{\partial J(n)} &=& \frac{1}{2} [\hdklmmn + \hdkmlmn]
\eea
then from \eref{eqn:gamma}
\bea
\gamma_{ij}(n) &=& \frac{1}{2} \delta_{ij} \lx[ \sum_m (\hdkimmn + \hdkmimn) \rx]
- \frac{1}{2} (1 - \delta_{ij}) (\hdkijmn + \hdkjimn).
\eea
Similarity, using the expression for the anisotropic $J_{ij}$
we obtain $\gamma^{L,T}_{ij}(n)$ for the anisotropic case,
\bea
\gamma^{L,T}_{ij}(n) &=& \frac{1}{2} \delta_{ij} \lx[ \sum_m (\hdkimmn + \hdkmimn) \Theta^{L,T}_{im}\rx] 
- \frac{1}{2} (1 - \delta_{ij}) (\hdkijmn + \hdkjimn) \Theta^{L,T}_{ij}.
\eea
Now, using \eref{eqn:deval}, it becomes easy to calculate the derivatives 
of the log-likelihood (\eref{eqn:loglikelihoodfull}) w.r.t each of its 
variable $J(n)$ and imposing its maximization we obtain,
\bea
1 - \CorrNexp &=& \frac{1}{N \Delta n} \sum_{k>1} \frac{1}{a_k} \frac{\partial a_k}{\partial J(n)} 
= \frac{\text{Tr}[A^{-1}\gamma(n)]}{N \Delta n}.
\label{eqn:dlogli1}
\eea
Similarly, for the anisotropic case the maximization 
of \eref{eqn:loglikelihoodanis} gives,
\bea
1 - \CorrNLTexp &=& \frac{1}{p^{L,T}(n) N \Delta n } \sum_{k>1} \frac{1}{a_k} \frac{\partial a_k}{\partial J^{L,T}(n)} 
= \frac{\text{Tr}[A^{-1}\gamma^{L,T}(n)]}{p^{L,T}(n) N \Delta n}.
\label{eqn:dlogli2}
\eea

\section{Numerical maximization of the log-likelihood}
\label{app:num}

The analytical expressions of the partition functions and its derivatives 
are not enough to analytically optimize the log-likelihood (\eref{eqn:loglikelihoodfull1}, \eref{eqn:loglikelihoodanis}). 
The reason is the following: the partition function 
and its derivatives are functions of the eigenvalues and eigenvectors of the 
network matrix $A_{ij}$ and it is not possible to diagonalize such
$N\times N$ matrix ($N$ is the number of birds of the flock) without the help of any numerical method.

There are two ways to approach such problem: (i) 
without providing the analytical expressions of the derivatives;
(ii) providing the analytical expressions of the derivatives. 
In the first case the  number of iteration needed for the optimization is much larger than in the second case, as we 
provide less information. 
Furthermore, as the dimension of the log-likelihood function 
gets larger, the number of iterations in the first case increases very rapidly.  
During each iteration step the most time-consuming part is the diagonalization of 
the matrix $A_{ij}$. Because the number of iterations is significantly smaller with method (ii)
than with (i), the whole computation gets much more efficient.
Practically speaking, for the largest flocks method (ii) is 
more than $10$ times faster than (i).
Therefore, the analytical expressions of the derivatives that we have (painfully) worked out in the previous sections 
are very useful to obtain a stable numerical solution in an efficient way. 

We use the minimizing routine {\sl gsl multimin fminimizer nmsimplex2}, 
belonging to the gnu scientific library \cite{Gsl}.
This optimization algorithm is based on 
Broyden-Fletcher-Goldfarb-Shanno (BFGS) algorithm \cite{multimin}. 
We provide as an input of the routines the analytical expressions of $Z$ 
and of the derivatives of $Z$ with respect to $J(n)$ (or to $J^L(n)$ and $J^T(n)$). 
For the diagonalization, we use the {\sl gsl eigen symmv} belonging to the gnu scientific library \cite{Gsl}.

In \fref{fig:iteration}, we plot the behavior of the parameter $J(1)$ 
and of the log-likelihood vs the iteration time, for three different initial conditions of $J(1)$.
It is clear that the solution is very stable and it is also reached very quickly.

\section{Data set}
\label{app:data}

\begin{table}[t]
\begin{tabular}{c|c|c|c|c|c|c}
\hline
\hline
EVENT & $N$ & $\Phi$  & $L$ ($\rm m$) & $n_c^\mathrm{exp} $ &
$n_c^\mathrm{step}$ & $r_c^\mathrm{exp} ($\rm m$) $\\
\hline
\hline
21--06 & 717  & 0.973   & 32.1  & 7.41  & 11.73 & 2.00 \\
\hline
25--10 & 1047 & 0.991   & 33.5  & 9.56  & 14.30 & 1.93 \\
\hline
25--11 & 1176 & 0.959   & 43.3  & 12.01 & 15.03 & 1.99 \\
\hline
28--10 & 1246 & 0.982   & 36.5  & 4.92  & 10.21 & 1.27 \\
\hline
29--03 & 440  & 0.963   & 37.1  & 4.46  & 7.67  & 1.94 \\
\hline
31--01 & 2126 & 0.844   & 76.8  & 6.11  & 12.37 & 2.97 \\
\hline
32--06 & 809  & 0.981   & 22.2  & 7.43  & 12.50 & 1.39 \\
\hline
42--03 & 431  & 0.979   & 29.9  & 7.79  & 14.60 & 2.08 \\
\hline
49--05 & 797  & 0.995   & 19.2  & 6.18  & 11.25 & 1.24 \\
\hline
57--03 & 3242 & 0.978   & 85.7  & 8.51  & 14.19 & 2.67 \\
\hline
58--06 & 442  & 0.984   & 23.1  & 7.39  & 12.89 & 1.63 \\
\hline
58--07 & 554  & 0.977   & 19.1  & 7.23  & 13.79 & 1.63 \\
\hline
63--05 & 890  & 0.978   & 52.9  & 5.26  & 10.21 & 1.98 \\
\hline
69--09 & 239  & 0.985   & 17.1  & 10.56 & 16.91 & 1.92 \\
\hline
69--10 & 1129 & 0.987   & 47.3  & 9.11  & 15.30 & 2.39 \\
\hline
69--19 & 803  & 0.975   & 26.4  & 14.76 & 21.56 & 1.97 \\
\hline
72--02 & 122  & 0.992   & 10.6  & 8.62  & 11.37 & 1.32 \\
\hline
77--07 & 186  & 0.978   & 9.1   & 5.97  & 12.36 & 1.17 \\
\hline
20111125-2   & 505  & 0.972  & 34.4  & 11.31 & 16.36 & 1.84 \\
\hline
20111214-4-1 & 139  & 0.985  & 32.8  & 5.24  & 9.97  & 1.64 \\
\hline
20111214-4-2 & 156  & 0.983  & 31.5  & 8.13  & 10.23 & 2.52 \\
\hline
20111215-1   & 394  & 0.994  & 49.8  & 8.48  & 20.62 & 1.68 \\
\hline
\hline
\end{tabular}
\caption{{\bf Flocks Data}:
Each line represents a different flocking event.
$N$ is the number of individuals in the flock,
$\Phi$ the average polarization,
$L$ the size of the flock (maximum distance between two birds),
$n_c^\mathrm{exp}$ the exponential decay range computed in this work and
$n_c^\mathrm{step}$ the interaction range of the step model of \cite{Bialek12}. Finally, $r_c^\mathrm{exp}$ represents the typical metric distance corresponding to the topological distance $n_c^\mathrm{exp}$: it can be seen that it is always much smaller than the flock's size.}
\label{table}
\end{table}

Experimental data were obtained from field observations on large flocks of starlings, ({\it Sturnus vulgaris}), in the field. Three dimensional trajectories of positions and velocities of each bird are obtained using stereometric photography and computer vision techniques \cite{Cavagna08a,Cavagna08b,Ballerini08a,Ballerini08b, Cavagna10,Attanasi13a,Attanasi13b}.  
As summarized in Table \ref{table}, we have analyzed 22 distinct flocking events, with sizes ranging from 122 to 3242 individuals and linear extensions from 9.1 to 85.7 m. All these events belong to two different sets. The first set  (events from 21-06 to 77-07 in Table \ref{table}) was taken in the period  2005-2008, with cameras shooting at $10$ frames-per-second (fps). \cite{Ballerini08a,Ballerini08b}. The second set (last 4 events in Table \ref{table}) was collected in the period between 2010-2012, with cameras shooting at $170$fps \cite{Attanasi13a,Attanasi13b}. All the events correspond to strongly ordered flocks, with polarization  between $\Phi = 0.844$ and $\Phi = 0.995$, hence justifying the spin wave expansion.
The duration of the observed events is on average 6 seconds and it ranges between 2.8 and 11.6 seconds.
The number of frames varies between 14 and 58 frames per event (with mean 30). These scales are set by experimental constraints. In a stereoscopic experiment a flocking event is filmed by several machine vision cameras located at different positions. To reconstruct the individual 3D trajectories the flock must be in the common field of view of all the cameras:  given the flock's typical distance from the apparatus (100-300 m), after 10-12 seconds at most the flock is out of the field of view (this time being shorter the larger/closer is the flock). Besides, the amount of digital information per second that can be grabbed by a high resolution stereo set-up is limited, which also sets a constraint on the amount of consecutive digital images that can be retrieved at high frequency. We note that these time durations represent significant scales in terms of the collective motion of natural flocks: starlings fly at approximately  $10$ m/s and a flock of thousands birds can perform a collective turn (global change of direction) in just a few seconds \cite{Attanasi13a}.


\begin{thebibliography}{99}
\bibitem{footnote1}
Note that the topological distance, $n$, is a discrete variable, hence the correlation function $\hat C(n)$ is in fact a discrete set of values and it should be indicated as $\hat C_n$. However, we prefer the $\hat C(n)$ notation, as it is reminiscent of the standard metric definition of
correlation function, $\hat C(r)$. For the same reason we inappropriately use Dirac's notation $\delta(k_{ij}-n)$ for what is actually a Kronecker's $\delta_{k_{ij},n}$.
\bibitem{footnote2}
Indeed the same calculation performed in Sec. \ref{sec:maxent} can be repeated in a metric fashion (using the metric distance, $r$, rather than the topological distance, $n$); in this case, we find that the metric range $r_c$ scales with the density, thus confirming the result of \cite{Ballerini08a} that the interaction in starling flocks is based on topological distance.
\end{thebibliography}
\end{document}